\documentclass[superscriptaddress, twocolumn, amsmath, amssymb, aps,
pra, letterdraft, notitlepage]{revtex4-1} %preview twocolumn

\usepackage{graphicx,graphics,epsfig,subfigure,times,bm,bbm,amssymb,amsmath,amsfonts,amsthm,mathrsfs,MnSymbol}
\usepackage[matrix,frame,arrow]{xypic}
\usepackage[normalem]{ulem}
\usepackage{slashed}
\usepackage{color}
\usepackage[usenames,dvipsnames,svgnames,table]{xcolor}
\usepackage[english]{babel}
\definecolor{darkblue}{rgb}{0.0,0.0,0.3}
\usepackage[colorlinks=true,
            linkcolor=red,
            urlcolor= darkblue,
            citecolor=blue]{hyperref}

\newcommand{\bes} {\begin{subequations}}
\newcommand{\ees} {\end{subequations}}
\newcommand{\bea} {\begin{eqnarray}}
\newcommand{\eea} {\end{eqnarray}}

\def\>{\rangle}
\def\<{\langle}

\newcommand{\bra}[1]{\langle#1|}
\newcommand{\ket}[1]{|#1\rangle}

\begin{document}

\title{Non-Markovian Noise Characterization with the Transfer Tensor Method}

\author{Yu-Qin Chen}
\affiliation{Tencent Quantum Laboratory, Tencent, Shenzhen, Guangdong, China, 518057}
\affiliation{International Center for Quantum Materials and School of Physics, Peking University, Beijing, China,100871}
\author{Kai-Li Ma}
%\affiliation{Tencent Quantum Laboratory, Tencent, Shenzhen, Guangdong, China, 518057}
\affiliation{The Chinese University of Hong Kong, Hong Kong}
\author{Yi-Cong Zheng}
\affiliation{Tencent Quantum Laboratory, Tencent, Shenzhen, Guangdong, China, 518057}
\author{Jonathan Allcock}
\affiliation{Tencent Quantum Laboratory, Tencent, Shenzhen, Guangdong, China, 518057}
\author{Shengyu Zhang}
\affiliation{Tencent Quantum Laboratory, Tencent, Shenzhen, Guangdong, China, 518057}
\affiliation{The Chinese University of Hong Kong, Hong Kong}
\author{Chang-Yu Hsieh}
\email{kimhsieh@tencent.com}
\affiliation{Tencent Quantum Laboratory, Tencent, Shenzhen, Guangdong, China, 518057}

\begin{abstract}
  We propose simple protocols for performing quantum noise spectroscopy based on the method of transfer tensor maps (TTM), [Phys. Rev. Lett. 112, 110401 (2014)].  The TTM approach is a systematic way to deduce the memory kernel of a time-nonlocal quantum master equation via quantum process tomography. With access to the memory kernel it is possible to (1) assess the non-Markovianity of a quantum process, (2) reconstruct the noise spectral density beyond pure dephasing models, and (3) investigate collective decoherence in multi-qubit devices.
 We illustrate the usefulness of TTM spectroscopy on the IBM \emph{Quantum Experience} platform, and demonstrate that the qubits in the IBM device are subject to mild non-Markovian dissipation with spatial correlations.
\end{abstract}

\maketitle

\section{Introduction}
Quantum information processing (QIP) is attracting increased attention from industry. As we enter the noisy intermediate-scale quantum
(NISQ) era~\cite{preskill_qm_18}, we face the near-term prospect of demonstrating non-trivial computations on quantum circuits consisting of 50-100 qubits without error correction. However, to sustain this level of engagement it is crucial to further improve the scale and achievable circuit depths of NISQ devices. At the core of this challenge is understanding the noise present in existing devices so that decoherence can be more effectively tamed \cite{breuer_theory_2007,de_vega_dynamics_2015,breuer_colloquium:_2016}. In this work, we propose a method for quantum noise spectroscopy based on the tomographically reconstructed transfer tensor maps (TTMs) \cite{PhysRevLett.112.110401} used in analyzing dissipative quantum dynamics. 
We demonstrate the utility of TTM noise spectroscopy by studying both simple theoretical models as well as IBM's \textit{Quantum Experience} (IBM Q) hardware on the cloud. 

While quantum process tomography (QPT) \cite{chuang_prescription_1997,poyatos_complete_1997,nielsen_chuang_book00} has been widely adopted for experimentally characterizing quantum-gate fidelities \cite{bialczak_quantum_2010,yamamoto_quantum_2010,rodionov_dnrey_prb2014} and environmental influences \cite{joelpnas11,howard_quantum_2006},
earlier theoretical and experimental efforts \cite{kofman_two-qubit_2009,howard_quantum_2006} have largely been confined to analyzing noise under the Markovian assumption.
Transfer tensor maps allow noise sources to be a comprehensively characterized by connecting
experimentally deduced QPT data with the time-nonlocal quantum master equation (TNQME) \cite{breuer_theory_2007} valid for general open quantum systems,

\begin{eqnarray} \label{eq:nz-mastereq}
\frac{d\rho(t)}{dt} = \mathcal{L}_s \rho(t) + \int^t_0 ds \mathcal{K}(t-s)\rho(s),
\end{eqnarray}
where $\rho(t)$ denotes the system's reduced density matrix,
$\mathcal{L}_s$ is the Liouville superoperator corresponding to the system Hamiltonian, and $\mathcal{K}(t)$ is the memory kernel that fully encodes environment-induced decoherence effects.
Through the tomographically reconstructed TTMs and translated memory kernel, experimentalists can more easily (1) characterize the non-Markovianity of a quantum process, (2) determine the noise power spectrum, and (3) estimate spatial and temporal correlations of nonlocal noise in multi-qubit systems.  Furthermore, the TTM method facilitates noise characterization beyond pure dephasing models and works without having to implicitly assume the quantum or classical nature of the noise.

While the TTM method \cite{PhysRevLett.112.110401,buser_initial_2017,kananenka_accurate_2016,gelzinis_applicability_2017,pollock_tomographically_2018} can model arbitrary noise, we will largely be concerned with stationary Gaussian noise, which can be fully characterized by two-time correlation functions and noise spectral density.
With the exception of a few special cases such as highly anharmonic spin baths \cite{prokof_stamp_rpp00,ma_liu_prb15}, this is a reasonable assumption. For instance, a spin bath surrounding a gated semiconductor quantum dot \cite{Hsieh_Shim_RPP201} can often be
modelled as classical Gaussian noise \cite{Hsieh_Cao_JCP2018,merkulov_efros_prb02,erlingsson_hyperfine_2002,witzel_sarma_prb14}. Also, the noise we observe often represents the overall effect of a system's interactions with many background excitations. At the level of interacting with a single excitation, the coupling strength is often very weak. However, the cumulative effect (due to many background excitations) of the noise can be strong and may dominate the open system dynamics. In this scenario \cite{makri_jpcb1999,caldeira_leggett_annals83}, our assumptions on the noise properties should hold.

It is instructive to compare TTM noise spectroscopy to another popular approach based on dynamical decoupling (DD) \cite{PhysRevA.58.2733,khodjasteh_lidar_prl05,khodjasteh_lidar_prl10,yang_liu_frontphys11,szankowski_environmental_2017}, which is an open-loop quantum control technique using $\pi$-pulses (at a frequency much higher than the characteristic time scale of the noise) in rapid succession such that qubits become effectively isolated from their environment. While DD spectroscopy \cite{alvarez_measuring_2011,zwick_maximizing_2016,norris_qubit_2016,paz-silva_multiqubit_2017} has recently been widely adopted to acquire noise statistics for quantum devices, results have largely been restricted to spin-based quantum circuits as the spectroscopy is based on the assumption that pure dephasing is the dominant decoherence mechanism \cite{alvarez_measuring_2011,zwick_maximizing_2016,norris_qubit_2016,paz-silva_multiqubit_2017,krzywda_dynamical-decoupling-based_2018,cywinski_dynamical_2014,yuge_measurement_2011,ma_proposal_2016}. This assumption is incompatible with certain types of physical qubits, such as  superconducting qubits which have extremely long coherence times \cite{hanhee_schuster_prl11,rigetti_gambetta_prb12,chen2014qubit} in comparison to the gating times and which have comparable $T_1$ and $T_2$ time scales. Secondly, the repeated use of $\pi$-pulse can potentially introduce adversarial effects\cite{shiokawa_lidar_pra04} if the pulses are not in actuality fast relative to the cut-off frequency of the noise spectral density. We shall demonstrate that the TTM approach can not only reconstruct the noise power spectrum beyond pure dephasing models but can also achieve noise characterization without the need for a series of $\pi$-pulses.

In multi-qubit circuits additional complexity emerges. As
most fault-tolerant schemes \cite{lidar_quantum_2013} assume independent noise  %aharonov:1997:176, 
on each qubit, it is crucial to have a quantitative diagnosis of the spatial and temporal correlations of decoherence
effects among qubits in the system.   DD noise spectroscopy \cite{zwick_maximizing_2016,norris_qubit_2016,paz-silva_multiqubit_2017,krzywda_dynamical-decoupling-based_2018} has been extended to use
multi-qubit registers as a probe to analyze spatial correlations of noise.  Similar to the
single qubit case, DD based correlated noise spectroscopy still primarily focuses on pure-dephasing models. Again, the advantages of single qubit TTM spectroscopy extends to the case of multi-qubit systems. 

In principle, TTM spectroscopy can be systematically applied to an arbitrary number of qubits. However, due to experimental constraints of performing high-dimensional QPT, it is most feasible to consider TTM based noise characterization for a small number of entangled qubits. In particular, we will only discuss two-qubit TTM spectroscopy in this work. However, recent advances in adopting techniques such as compressed sensing \cite{gross_liu_prl10,riofrio_gross_natcomm17} and variational ans\"atze derived from machine learning \cite{torlai_mazzola_natphys18,rochcetto_grant_npjqip18,giacomo_melko_arxiv18} and matrix product states \cite{cramer_plenio_nat2010,lanyon_maier_natphys17} have greatly improved the prospects of tomographic scalability. Additional efforts, such as using machine learning \cite{palmieri_kovlakov_arxiv19} to tame state preparation and measurement errors, without resorting to expensive gate set tomography and similar techniques \cite{seth_gambetta_pra13,kohout_gamble_natcomm13}, could further strengthen the attractiveness of TTM noise spectroscopy in comparison to other well-established approaches in the near future.

The remainder of this work is organized as follows. Sec.~\ref{sec:ttm} contains an introduction to transfer tensor maps. Sec.~\ref{sec:ttm_spect} discusses the proposed TTM noise spectroscopy. Sec.~\ref{sec:ttm_theoretical} describes the usage of
TTM noise spectroscopy with examples based on theoretical models.
Sec.~\ref{sec:ibm} summarizes a testing of the TTM method on IBM's \textit{Quantum Experience}
hardware.  Section VI concludes.

%%%%%%%%%%%%%%%%%%%%%%%%%%%%%%%%%%%%%%%%
\section{Transfer tensor maps}\label{sec:ttm}
%%%%%%%%%%%%%%%%%%%%%%%%%%%%%%%%%%%%%%%%

We first recall the essential theoretical background of TTM as originally formulated in Ref.~\onlinecite{PhysRevLett.112.110401}.
In this work, we are concerned with a collection of qubits governed by the following Hamiltonian,
\begin{eqnarray}\label{eq:H1}
H(t) & = & H_s + H_{sb}(t) \nonumber \\
& = & H_s + \sum_{i,\alpha} g_{i} B^\alpha_i(t) \sigma^\alpha_i,
\end{eqnarray}
where $H_s$ is a time-independent system Hamiltonian for the qubits, and $H_{sb}$ describes the system-noise
coupling. $\sigma_i^\alpha$ is a Pauli operator with the index $i$ labeling the qubits and the index $\alpha$ one of the $\{x,y,z\}$ Cartesian components. $B^\alpha_i(t)$ is a bath operator explained below.
Let us clarify the distinction between quantum and classical noise defined in this work. When the environment consists of quantum systems, $B^\alpha_i(t) = e^{-i H_b t}B^\alpha_i(0)e^{i H_b t}$ is a bath operator in the interaction picture with
respect to $H_b$, the environment Hamiltonian.
%Commonly,this quantum environment could be effectively approximated with a collection of harmonic oscillators.
When the environment consists of classical stochastic baths, $B^\alpha_i(t)$ are real-valued stochastic processes. In both cases, we only consider Gaussian noise, which is fully characterized by the first two statistical moments $\langle B^\alpha_i(t) \rangle$ and $\langle B^\alpha_i(t) B^\beta_j(t') \rangle$. % = \xi_{ij}(t-t')$.

Furthermore, we assume an initial state $\rho(0)$ of the qubits to be independent of the environment.
Time evolution for an open quantum system can be cast in the following form,
\begin{eqnarray}\label{eq:dynmap}
    \rho(t) & = & \left \langle \exp_+\left(-i\int^t_0 ds H(s)\right) \rho(0) \exp_-\left(i\int^t_0 ds H(s)\right) \right \rangle \nonumber \\
&=& \mathcal{E}(t)\rho(0),
\end{eqnarray}
where the $\pm$ subscript on the exponential functions denotes the (anti-)chronological time ordering of the time-evolution operator and $\mathcal{E}(t)$ is the dynamical map relating the time-evolved reduced density matrix back to the initial state. The bracket $\langle \cdots \rangle $ denotes an average over the environmental degrees of freedom.  Experimentally, the dynamical maps $\mathcal{E}_n$ for a $d$-level quantum state are derived from an ensemble of QPTs obtained under $d^2$ different initial conditions.

The QPTs are performed at equidistant time intervals, i.e. $t_k = k \delta t$, and we may write $\mathcal{E}_k \equiv \mathcal{E}(t_k)$.
Transfer tensors are then defined by
\begin{eqnarray}\label{eq:ttm}
T_n \equiv \mathcal{E}_n - \sum_{m=1}^{n-1} T_{n-m}\mathcal{E}_m,
\end{eqnarray}
with $T_1 = \mathcal{E}_1$.
Replacing the dynamical maps $\mathcal{E}_n$ in Eq.~(\ref{eq:dynmap}) with $T_n$ in Eq.~(\ref{eq:ttm}), one obtains
\begin{eqnarray}\label{eq:ttm2}
\rho(t_n) = \sum_{m=1}^{n-1} T_m \rho(t_{n-m}).
\end{eqnarray}
Equation (\ref{eq:ttm2}) shows that the dynamical evolution up to time $t_n$  depends on the system's history in the presence of noise. The TTM $T_m$ encodes how much the system's state at an earlier time $t_{n-m}$ contributes to the formation of the current state $\rho(t_n)$.
A time translational invariance is implied for the transfer tensor maps in Eq.~(\ref{eq:ttm2}) as they are labeled by the time difference $m\delta t$
between the two density matrices $\rho(t_n)$ and $\rho(t_{n-m})$. Equation (\ref{eq:ttm2}) holds as we only consider time-independent $H_s$.
In general, the dynamical evolution should only significantly depend on the system's history up to certain point. This implies that one may accurately estimate the quantum dissipative dynamics when the the summation in Eq.~(\ref{eq:ttm2}) is truncated at large enough index $K$, $i.e.$ drop all summands for time $t \geq t_K$. This theoretical insight has an important consequence for experiment: \textit{one only needs to perform QPTs up to time $t_K = K\delta t$. Beyond $t_K$, all quantum dynamical information can be recursively determined via Eq.~(\ref{eq:ttm2})}. 

\section{TTM based Noise Spectroscopy}\label{sec:ttm_spect}
%%%%%%%%%%%%%%%%%%%%%%%%%%%%%%%%%%%%%%%%

When the time increment $\delta t$ is small, Eq.~(\ref{eq:ttm2}) essentially prescribes a numerical solution to Eq.~(\ref{eq:nz-mastereq}) with the identification
\begin{eqnarray}\label{eq:ttm-mem}
T_n = (1+\mathcal{L}_s \delta t)\delta_{n,1} + \mathcal{K}(t_n)\delta t^2,
\end{eqnarray}
%where $\mathcal{K}_n = \mathcal{K}(n \delta t)$.
with $\delta_{n,1}$ the Kronecker delta function. The translated memory kernel of the TNQME, Eq.~(\ref{eq:nz-mastereq}), is the foundation of the noise spectroscopy method to be explicated in this section.

%%%%%%%%%%%%%%%%%%%%%%%%%%%%%%%%%%%%%%%%
\subsection{A measure of non-Markovianity }\label{sec:measure}
%%%%%%%%%%%%%%%%%%%%%%%%%%%%%%%%%%%%%%
Transfer tensor maps can be used to gauge the non-Markovianity of a quantum dynamical process. A simple argument is as follows \cite{PhysRevLett.112.110401}. A Markovian process requires only the present state to determine the next state, i.e. $\rho(t_n) = T_1 \rho(t_{n-1})$ for all $t_n$. Hence, there should be only one non-trivial transfer tensor $T_1$ (i.e. the Frobenius norm $\vert T_n \vert = 0$ for $n \geq 2$) for any  $\delta t$. Whenever multiple transfer tensors are needed to accurately approximate Eq.~(\ref{eq:ttm2}) the environment can be qualitatively argued as being non-Markovian. Alternatively, recall that the transfer tensors $T_{n}$ (stored as matrices) encode the memory kernel $\mathcal{K}_n$ for $n \geq 2$ in Eq.~(\ref{eq:ttm-mem}) when $\delta t$ is small. The temporal profile of the memory kernel, another qualitative indicator of non-Markovianity, can be gauged by the norm of TTM matrices constructed at different time points. %In this study, we adopt the Frobenius norm throughout this study. 

Counting the number of TTMs with sizable norm, however, does not fully capture the multifaceted nature of a non-Markvoian quantum process \cite{breuer_colloquium:_2016,rivas_quantum_2014,pollock_non-markovian_2018}. Transfer tensor maps may be used though, in combination with advanced theory of open quantum systems, to construct a variety of more insightful non-Markovianity measures. As an example, consider the Bloch volume measure \cite{lorenzo_geometrical_2013} put forward by Lorenzo, Plastina and Paternostro.
This scheme uses the Bloch sphere representation of a qubit as
\begin{eqnarray}
\rho(t) = \frac{1}{2} + \sum_\alpha r^\alpha(t) \sigma^\alpha,
\end{eqnarray}
with the Bloch vector $\mathbf{r}(t)=(r^x,r^y,r^z)$. Each Bloch vector component is determined via $r^\alpha(t)=\text{Tr}(\sigma^\alpha \rho(t))$. In this geometric view of quantum dynamics, the action of
$\mathcal{E}_n$ on $\rho(0)$ can be represented as an affine transformation on $\mathbf{r}$, i.e. $\mathbf{r}_n = \mathcal{M}_n \mathbf{r}_0 + \mathbf{c}_n$.
It is known that the volume $V(t_n)=\text{det}\vert \mathcal{M}_n \vert$ of accessible states at time $t_n$ is a monotonically decreasing function for Markovian dynamics. The Bloch volume measure $\mathcal{N}_V$ is then defined as
\begin{eqnarray}\label{eq:bvm}
\mathcal{N}_V=\frac{1}{V(0)} \int_{\partial_t V(t) > 0} \partial_t V(t),
\end{eqnarray}
which accumulates the positive rate of change when the volume of accessible states increases due to deviations from Markovianity.
Note that $\mathcal{N}_V$ is not a comprehensive non-Markovian measure as it is insensitive to non-Markovianity associated with the $\mathbf{c}_n$ vector. Nevertheless, it is one of the simplest measures to evaluate that still gives insight into the underlying dynamical process.

A few toy models aside, evaluating
Eq.~(\ref{eq:bvm}) requires tracking the time-dependent dynamical maps in order to observe the temporary reversal of the volume contractions and information backflow for non-Markovian processes in an experimental setting. By a straightforward modification of Eq.~(\ref{eq:ttm2}),
we arrive at a recursive relation $\mathcal{E}_n = \sum_{m=1}^K T_m \mathcal{E}_{n-m}$. This relation significantly reduces the number of required QPTs when one attempts to experimentally evaluate Eq.~(\ref{eq:bvm}), because one can re-use a small number of QPTs (up to time $t_K$) to numerically generate addition dynamical maps at $t > t_K$.

%%%%%%%%%%%%%%%%%%%%%%%%%%%%%%%%%%%%%%%%
\subsection{Noise spectral density: weak noise-coupling}\label{sec:specden}
%%%%%%%%%%%%%%%%%%%%%%%%%%%%%%%%%%%%%%%%
For simplicity, we explain how to construct the noise spectral density for a single qubit coupled to quantum noise. Extension of the this method to the two-qubit case is straightforward. Furthermore, in App.~\ref{app:proj} we show that the the case of classical noise may also be dealt with using only minor modifications. The weak-coupling (to noise sources) implies that it is sufficient to approximate the memory kernel with the leading order term of a perturbative expansion with respect to $H_{sb}$.

The exact memory kernel for an arbitrary quantum bath can be written as follows,
\begin{eqnarray}\label{eq:memker}
\mathcal{K}(t, t^\prime) & = & \mathcal{P}\mathcal{L}(t)\exp_+\left[\int^t_{t^\prime} ds \mathcal{QL}(s)\right]\mathcal{QL}(t^\prime) \mathcal{P},
\end{eqnarray}
where projection operators $\mathcal{P}$ and $\mathcal{Q}=1-\mathcal{P}$ are defined by $\mathcal{P}\Omega(t)=\text{Tr}_b(\Omega(t))\otimes\rho_b$, i.e. $\mathcal{P}$ projects a system-bath entangled quantum state $\Omega(t)$
to a factorized form consisting of a system part $\rho(t)=\text{Tr}_b \Omega(t)$, and a bath part $\rho_b$ which is a stationary state with respect to the bath Hamiltonian $H_b$ defined in Eq.~(\ref{eq:H1}). We note that the kernel is a stationary process $\mathcal{K}(t,t^\prime)=\mathcal{K}(t-t^\prime)$ when the noise satisfies the stationary Gaussian conditions.

We first discuss TTM noise spectroscopy in the weak noise-coupling regime, which may be a reasonable approximation for high quality qubits well protected from noise, using a second-order perturbative treatment of the memory kernel. The general case will follow in the next section.

The memory kernel for  Gaussian noise can be expressed as $\mathcal{K}(t)=\sum_{n=1}^\infty \mathcal{K}_{2n}(t)$, where $\mathcal{K}_{2n}(t)$ corresponds to the order $2n$ expansion of the Hamiltonian with respect to $H_{sb}$.
Keeping only the leading (i.e. second order) term  gives
\begin{eqnarray}\label{eq:corr}
 & & \mathcal{K}_{2}(t)(\boldsymbol{\cdot})  \approx \left \langle \mathcal{L}_{sb}(t)
\mathcal{L}_{sb}(0) \right \rangle ( \boldsymbol{\cdot} ) \\
%& = & \text{Tr}_b\left(\rho_b \mathcal{L}_{sb}(t) \mathcal{L}_{sb}(t^\prime)\right) (\boldsymbol{\cdot} ) \nonumber \\
& & \,\,\, = \sum_{\alpha\alpha^\prime} [\sigma^\alpha, C_{\alpha\alpha^\prime}(t) \sigma^{\alpha^\prime}(t) (\boldsymbol{\cdot} )
%\nonumber \\ & &
- C^*_{\alpha\alpha^\prime}(t)(\boldsymbol{\cdot}) \sigma^{\alpha^\prime}(t) ], \nonumber
%& = & \left\langle\mathcal{L}_1(t) \mathcal{L}_1(t^\prime)\right\rangle + \left\langle\mathcal{L}_2(t) \mathcal{L}_2(t^\prime)\right\rangle
%+\left\langle\mathcal{L}_1(t) \mathcal{L}_2(t^\prime)\right\rangle + \left\langle\mathcal{L}_2(t) \mathcal{L}_1(t^\prime)\right\rangle
\end{eqnarray}
where $\sigma^{\alpha}(t)=\exp(-iH_{s}t)\sigma^\alpha \exp(iH_s t)$ and the bath correlation function is given by
\begin{eqnarray}
C_{\alpha\alpha^\prime}(t)= g^2 \left\langle \hat{B}^\alpha(t) \hat{B}^{\alpha^\prime}(0) \right\rangle.
\end{eqnarray}
We suppress the index $i$ on Pauli matrices and noise operators $\hat B^\alpha(t)$ as we only consider one qubit here.  Since the TTM matrices and the memory kernel are related via Eq.~(\ref{eq:ttm-mem}), every experimentally determined TTM matrix element is a linear combination of 9 correlation functions, $\{C_{\alpha\alpha'}(t)\}$, when the second-order perturbation, Eq.~(\ref{eq:corr}), holds. To numerically process the data and determine the correlation functions, we resort to solving a sequence of optimization problems,
\begin{eqnarray}\label{eq:ncorr}
  \underset{C_{\alpha\alpha^\prime}(t_n)}{\text{argmin}} & \,\, &
 \{  \vert \mathcal{K}_2(t_n; C_{\alpha\alpha^\prime}(t_n))  - \mathcal{K}_{\text{exp}}(t_n) \vert \\
 & + &  (1-\delta_{t_n,t_0}) \lambda_n \sum_{\alpha,\alpha^\prime} \vert C_{\alpha\alpha^\prime}(t_n)-C_{\alpha\alpha^\prime}(t_{n-1}) \vert \}. \nonumber
\end{eqnarray}
where $\mathcal{K}_{\text{exp}}$ is the experimentally determined memory kernel. We numerically identify a set of $\{C_{\alpha\alpha^\prime}(0)\}$ to minimize the difference between the theoretically constructed second-order memory kernel matrix and the experimentally obtained memory kernel matrix at $t=0$. Subsequently, we determine the numerical values of the correlation functions at $t=t_1, t_2 \dots$ by solving the same minimization problem with an additional regularization enforcing the continuity of the correlation functions with hyperparameters $\lambda_n$.

Once the noise correlation function is determined, the corresponding spectral density can be determined by invoking the fluctuation-dissipation theorem \cite{yan_xu_annphyschem05}, which gives
\begin{eqnarray} \label{eq:omega}
J_{\alpha\alpha^\prime}(\omega) &=& \frac{1}{2} \int^\infty_{-\infty} dt e^{i\omega t}\left[ C_{\alpha\alpha^\prime}(t)-C^*_{\alpha\alpha^\prime}(t) \right].
\end{eqnarray}
%The quantum spectral density obeys a simple relation $J_{\alpha\alpha^\prime}(\omega)=-J_{\alpha^\prime\alpha}(-\omega)$ with $J_{\alpha\alpha^\prime}(0)=0$.
%The positivity of $J_{\alpha\alpha^\prime}(\omega)$ for $\omega >0$ is ensured if $C_{\alpha\alpha^\prime}(\omega)$, the Fourier transform of $C_{\alpha\alpha^\prime}(t)$, is positive. The quantum spectral density obeys a simple relation $J_{\alpha\alpha^\prime}(\omega)=-J_{\alpha^\prime\alpha}(-\omega)$ with $J_{\alpha\alpha^\prime}(0)=0$.
The positivity of $J_{\alpha\alpha^\prime}(\omega)$ for $\omega >0$ is ensured if $C_{\alpha\alpha^\prime}(\omega)$, the Fourier transform of $C_{\alpha\alpha^\prime}(t)$, is positive.

Throughout this derivation, we do not make any assumption of a pure dephasing model. This makes TTM noise spectroscopy particularly suitable for superconducting qubits as further detailed in Sec.~\ref{sec:ibm}.

%%%%%%%%%%%%%%%%%%%%%%%%%%%%%%%%%%%%%%%%
\subsection{Noise spectral density: beyond weak coupling}\label{sec:specden2}
%%%%%%%%%%%%%%%%%%%%%%%%%%%%%%%%%%%%%%%
Beyond the weak-coupling limit, one must retain more terms in the expansion of the memory kernel, $\mathcal{K}(N,t)=\sum_{n=1}^N \mathcal{K}_{2n}(t)$. It is thus more tedious to infer the correlation functions from the TTM matrix elements. To overcome this challenge, we propose to tomographically reconstruct a family of $N$ memory kernels $\mathcal{K}_i(t)$ under different conditions. The $i$-th case is distinguished from other cases by $\omega_{s,i}$, the system's characteristic energy scale of $H_s = \omega_{s,i} \sigma^z$. After casting each $\mathcal{K}_i(t)$ in its characteristic time scale $\propto (1/\omega_{s,i})$, the second-order memory kernel for the original problem can be extracted from linear combinations of the scaled $\mathcal{K}_i(t)$, and one may follow the numerical recipe introduced in Sec.~\ref{sec:specden} to recover the correlation functions and the associated spectral density.

We rely on several assumptions. First, we take the system Hamiltonian to be $H_s = \omega_s \sigma^z$, a simple bias.  Second, the qubit's characteristic energy scale, $\omega_s$ can be freely adjusted over a certain range $[\omega_{min},\omega_{max}]$, which is a reasonable assumption for tunable quantum devices. Third, the adjustment of $\omega_s$ should not alter the noise source. These requirements are compatible with many physical systems. For instance, a recent experiment\cite{capellaro_nvcenter_noise_2018} on spin-based qubits in NV center fulfills these requirements. 

The procedure goes as follows. For a chosen energy bias $\omega_{s,i}$, the quantum dynamics is governed by the Hamiltonian $H=\omega_{s,i} \sigma^z + H_{sb}(t)$. If a time unit of $(1/\omega_{s,i})$ is adopted, then the dimensionless Hamiltonian reads $\tilde{H}=\sigma^z+g \tilde{H}_{sb}(t)$ with $g \propto 1/\omega_{s,i}$.  By performing $N$ such experiments with distinct energy biases $\omega_{s,i}$ with $i=0 \dots N-1$, one may construct a set of dimensionless memory kernels, which are related by
\begin{eqnarray}
\tilde{\mathcal{K}}_{2n,i}(t) =\gamma_i^{2n} \tilde{\mathcal{K}}_{2n,0}(t),
\end{eqnarray}
where $\gamma_i=\omega_{s,0}/\omega_{s,i}$ and $\tilde{\mathcal{K}}_{2n,i}$ denotes the $2n$-th order term of the
memory kernel for an energy bias $\omega_{s,i}$.
If we assign the original case of interest as the $i=0$ case, then we may define a matrix,
\begin{eqnarray}
\mathcal{A} = \left[\begin{array}{cccc}
1 & 1 & \cdots & 1 \\
\gamma_1^2 & \gamma_1^4 & \cdots & \gamma_1^{2N} \\
\vdots &\vdots & \cdots & \vdots \\
\gamma_{N-1}^2 & \gamma_{N-1}^4 & \cdots & \gamma_{N-1}^{2N}
\end{array}
\right],
\end{eqnarray}
which encodes the linear algebraic relation $\mathcal{A}[\tilde{\mathcal{K}}_{2,0} \cdots \tilde{\mathcal{K}}_{2N,0}]^T=
[\tilde{\mathcal{K}}(1) \cdots \tilde{\mathcal{K}}(N)]^T$.
By converting the matrix $\mathcal{A}$ into a lower triangular form, one
learns how to linearly combine $\tilde{\mathcal{K}}_{2n,i}(t)$ to recover $\tilde{\mathcal{K}}_{2,0}(t)$, which is accurate up to the $2N$-th order perturbative truncation of the memory kernel stated in the beginning of this section. Finally, $\mathcal{K}_{2,0}(t)$ is recovered when one reverts back to the original time unit scale.

%%%%%%%%%%%%%%%%%%%%%%%%%%%%%%%%%%%%%%%%
\subsection{Two-qubit environment spectroscopy}\label{sec:deltattm}
%%%%%%%%%%%%%%%%%%%%%%%%%%%%%%%%%%%%%%%%
In a two-qubit circuit, unintentional entanglement between the qubits may be built up when the qubits
share a common noise source. It is thus desirable to unravel the two-qubit TTMs into single-qubit and two-qubit contributions so that one may not only detect the presence of common noise but also determine a quantitative estimate of its contribution to collective deocherence.

First, we write two-qubit dynamical maps in the basis of tensor products of single-qubit basis. In the
such a basis, two-qubit dynamical maps may be rigorously decomposed as
\begin{eqnarray}
\mathcal{E}_n & \equiv & \mathcal{E}_{n,1} \otimes \mathcal{E}_{n,2} + \delta \mathcal{E}_n \nonumber \\
& = & \bar{\mathcal{E}}_n + \delta \mathcal{E}_n,
\end{eqnarray}
where $\mathcal{E}_{n,1}$  and $\mathcal{E}_{n,2}$ are the reduced dynamical maps after tracing out the system 2 and system 1, respectively.
Details of this decomposition of bipartite dynamical maps are given in App.~\ref{app:MtxRep}. If the two qubits are not directly coupled, then the dynamical maps can be exactly separated with $\delta \mathcal{E}_n = 0$ for all $n$.  In some earlier works \cite{kofman_two-qubit_2009}, $\delta \mathcal{E}_n$ is directly used to
diagnose collective decoherence.

The decomposition of $\mathcal{E}_n$ into a separable component $\bar{\mathcal{E}}_n$ and correlated component $\delta\mathcal{E}_n$ is particularly useful when second-order perturbation theory holds. This is assumed for our analysis of two-qubit systems subsequently discussed in Sec.~\ref{sec:toy2} and \ref{sec:ibm2}, respectively. Whenever the observed dynamical process falls beyond this perturbative regime, one may find $\vert \delta \mathcal{E}_n \vert$ to be comparable to or even dominate  $\vert \bar{\mathcal{E}}_n \vert$. This implies the physical picture of having two localized quantum subsystems is no longer an appropriate basis to discuss a strongly correlated quantum system.

A similar set of separable and correlated tensor maps may also be defined as
\begin{eqnarray}
\bar{T}_n & = &\bar{\mathcal{E}}_n - \sum_{m=1}^{n-1} \bar{T}_{n-m} \bar{\mathcal{E}}_{m}, \\
T_n & = & \bar{T}_n + \delta T_n,
\end{eqnarray}
with $\bar{T}_1 = \bar{\mathcal{E}}_1$, and the correlated memory kernel identified as
\begin{eqnarray}\label{eq:delta-Tn}
\delta T_n = \delta\mathcal{L}\delta t \delta_{n,1} + \delta\mathcal{K}_n\delta t^2.
\end{eqnarray}
The Liouville superoperator $\delta\mathcal{L}$ reveals whether a direct qubit-qubit coupling term $\delta H_{12}$, unrelated to the noise, is present in the qubit Hamiltonian $H_s$. It is beneficial to identify the main cause of collective decoherence, and devise a corresponding strategy to improve the coherence of the quantum hardware.
If one simply looks at the dynamical map $\delta \mathcal{E}_n$, it is hard to distinguish the source of collective decoherences: (1) a direct coupling $\delta H_{12}$, (2) genuine correlated noises, or (3) an interplay of both factors. To complicate the situation, two entangled qubits suffer collective decoherence even when they are subjected to independent noise with $\delta H_{12}=0$.

In pure dephasing models, it is possible to quantitatively distinguish between the contributions of the two factors.  An unintentional coupling $\delta H_{12}$ will give rise to the $\delta \mathcal{L}$ term in \eqref{eq:delta-Tn} only and nothing else. On the other hand, effects of correlated noise sources are exclusively encoded in the correlated part of the memory kernel. In more general decoherence models, both factors contribute to the correlated memory kernel. Nevertheless, $\delta \mathcal{L}$ still only encodes the effects given by $\delta H_{12}$. We may estimate the $\delta \mathcal{L}$ term by performing
short-time QPT at 2 different time step size $\delta t$ and $\delta t^\prime$. Linear combinations of the two different maps, $\mathcal{E}_1$ and $\mathcal{E}^\prime_1$, can be used to isolate the two terms as $\delta \mathcal{L}$ and $\delta \mathcal{K}$ scale as $\delta t$ and $\delta t^2$, respectively. This analysis therefore still provides a gauge on the relative importance of the two factors for collective decoherence.

%%%%%%%%%%%%%%%%%%%%%%%%%%%%%%%%%%%%%%%%%%%%%%%%%%%%%%%%%
\section{TTM spectroscopy on theoretical models}\label{sec:ttm_theoretical}
%%%%%%%%%%%%%%%%%%%%%%%%%%%%%%%%%%%%%%%%%%%%%%%%%%%%%%%%%
We first illustrate the usefulness of  TTM spectroscopy through numerical experiments, using real-valued Gaussian random noise. As we shall see, the numerically processed
TTM data reveals much about the noise properties.

%%%%%%%%%%%%%%%%%%%%%%%%%%%%%%%%%%%%%%%%%%%%%%%%%%%%%%%%%
\subsection{One-Qubit Case}\label{sec:toy1}
%%%%%%%%%%%%%%%%%%%%%%%%%%%%%%%%%%%%%%%%%%%%%%%%%%%%%%%%%
We first consider a one-qubit pure phasing model, which is commonly used to illustrate DD spectroscopy. The Hamiltonian reads $H_s=\omega_{s}\sigma^z$ and $H_{sb}=B^z(t)\sigma^z$. The noise $B^z(t)$ satisfy the statistical moments: $\langle B^z(t) \rangle = 0$ and $C_{zz}(t-t^\prime)=\langle B^z(t)B^z(t^\prime)\rangle = \lambda e^{-\left|t-t'\right|}\cos[\omega_{c}(t-t')]$. For these cosine functions modulated with an exponentially decaying envelope, the corresponding spectral density has the Lorentzian profile. %$S(\omega)=\lambda/(1+(\omega-\omega_c)^2)+\lambda /(1+(\omega+\omega_c)^2)$.

The pure dephasing dynamical maps are given by
\begin{eqnarray}\label{eq:toy-dmap}
\mathcal{E}_n = \left[\begin{array}{cccc}
1 & 0 & 0 & 0 \\
0 & e^{-\Gamma(t_n)+i2\omega_s t} & 0 & 0 \\
0 & 0 & e^{-\Gamma(t_n)-i2\omega_s t} & 0 \\
0 & 0 & 0 & 1 \end{array} \right],
\end{eqnarray}
with
\begin{eqnarray}\label{eq:gammat}
\Gamma(t)=\frac{4}{\pi}\int^\infty_0 d\omega\frac{S(\omega)}{\omega^2}[1-\cos(\omega t)].
\end{eqnarray}
Note that these dynamical maps satisfy  $\rho(t_n) = \mathcal{E}_n \rho(0)$ as required.
It is straightforward to verify that these maps are not divisible, i.e. $\mathcal{E}_{n+m}\neq \mathcal{E}_n\mathcal{E}_m$ for this
pure-dephasing model; hence, the dynamics is clearly non-Markovian.

\begin{figure}[htp]
\centering
\includegraphics[width=0.38\textwidth]{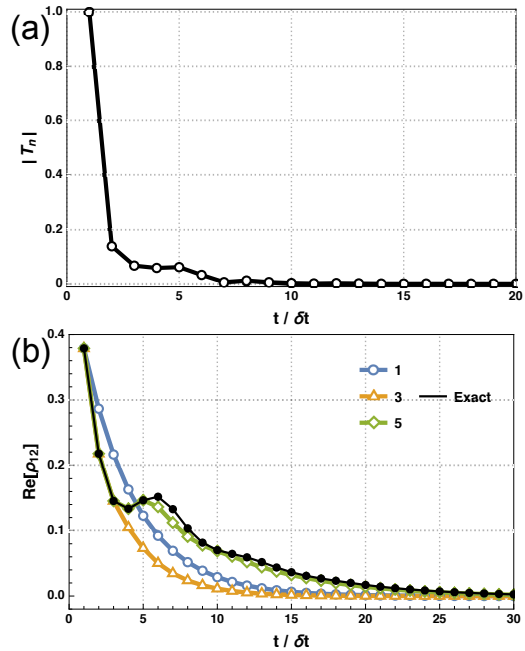}
\caption{Transfer tensor maps (TTM) for a single-qubit model. (Top): Frobenious norm distribution of the TTMs, $\vert T_n \vert$.  (Bottom): The real part of the off-diagonal matrix element of $\rho(t)$ with the initial condition  $(|0 \rangle+|1 \rangle)/\sqrt{2}$. The model parameters: $H_s=0.1\sigma_z$, $H_{sb}=B^z(t)\sigma ^z$, $C_{zz}(0)=\lambda=4$ and $\delta t=0.2$.}
\label{fig:toy-ttmnorm-predict}
\end{figure}

In the upper panel of Fig.~\ref{fig:toy-ttmnorm-predict} we plot the norms $\vert T_n \vert$ of different TTM matrices.
and note that more than one TTM matrix has sizable norm. This observation confirms the non-Markovianity of the dynamical process is positively correlated with TTM norm distribution. Furthermore, the norms of higher order matrices are increasingly suppressed, justifying the earlier claim that we should be able to truncate Eq.~(\ref{eq:ttm}) to include only the first few TTMs with non-trivial norms when we make dynamical predictions based on Eq.~(\ref{eq:ttm2}). In the lower panel of Fig.~\ref{fig:toy-ttmnorm-predict} we plot the norm of the off-diagonal matrix element, $\vert \rho_{12}(t) \vert$, as a function of time. The curve with black circles is the exact result $e^{-\Gamma(t)}$, which requires explicitly evaluating the integral in Eq.~(\ref{eq:gammat}) at every time point.  The other curves are results obtained by using different numbers of TTMs in the way prescribed by Eq.~(\ref{eq:ttm2}) to predict quantum dynamical evolution. In this case, we consider using the first $n=1,3,5$ TTMs, respectively.  Accurate results are obtained for the entire simulation duration when a sufficient number ($n=5$) of TTMs are taken into account.
\begin{figure}[htp]
\centering
\includegraphics[width=0.35\textwidth]{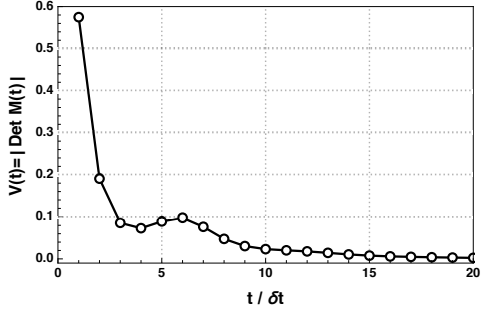}
\caption{The time-dependent Bloch volume $V(t)$ is plotted. The non-Markovianity is manifested by the temporary reversal of an otherwise monotonic decay profile of $V(t)$ around $t = 5 \delta t$.}
\label{fig:toy-measure}
\end{figure}

Next we analyze the non-Markovianity of the dynamical process by plotting the Bloch volume measure $ V(t) = \text{det} \vert M(t) \vert$  in Fig.~(\ref{fig:toy-measure}). One sees the signature of non-Markovianity manifested in the temporary increase of the Bloch volume. This increase happens around the time interval spanned by $t_4$, $t_5$ and $t_6$ in the figure.
The exact result and TTM prediction (based on first 5 data points) agree very well across the entire
simulation time window. This is an encouraging indication that TTM may facilitate the implementation of theoretical non-Markovianity measures in an experiment.

\begin{figure}
\centering
\includegraphics[width=0.4\textwidth]{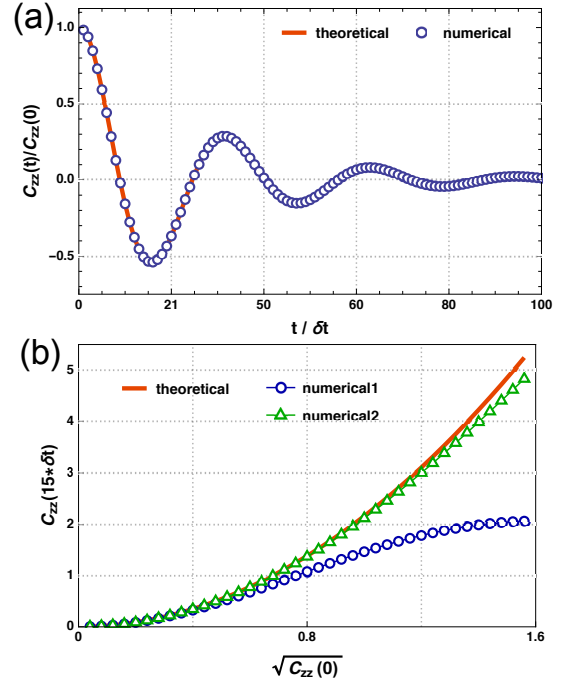}
\caption{TTM noise spectroscopy. (Top): Weak coupling case. Noise correlation function, $C_{zz}(t)$, is plotted as as function of time. The exact result is compared to the one extracted from the memory kernel under the assumption that $\mathcal{K}(t) \approx \mathcal{K}_2(t)$. Model parameters: $H_s=0.02\sigma_z$,  $C_{zz}(0)=\langle B^z(t) B^z(0)\rangle=0.01$ and $dt=0.04$.
(Bottom): Strong coupling case.  The noise correlation function at a specific time $C_{zz}(t=15\delta t)$ is plotted as a function of the system-noise coupling strength $C_{zz}(0)=\lambda$.  The first numerical result, $\mathcal{K}(t) \approx \mathcal{K}_2(t)$, deviates significantly from the exact result under the strong noise coupling limit, while the second numerical result, $\mathcal{K}(t)\approx\mathcal{K}_2(t)+\mathcal{K}_4(t)$, matches well with the exact one. Model parameters:  $H_s=0.02\sigma_z$,  $C_{zz}(0)\in{(0,1.6^2)}$ and $dt=0.04$. 
 }
\label{fig:toy-specden}
\end{figure}

A critical role of a quantum noise spectroscopy is determining the noise spectral density.  Since the spectral density is the Fourier transform of a correlation function, we will be content if we obtain the noise correlation function.  First, we consider the case of weakly coupled noise where the approximation
$\mathcal{K}(t)\approx \mathcal{K}_2(t)$ is valid. We use the numerical method, Eq.~(\ref{eq:ncorr}), proposed in Sec.~\ref{sec:specden} to infer the correlation function. In the upper panel of Fig.~(\ref{fig:toy-specden}), we plot the numerically recovered correlation function based on TTM data, and it agrees well with the theoretical correlation function that the random noise $B^z(t)$ must satisfy in our numerical experiment. We next explore a broader range of system-noise coupling strength. It is no longer valid to assume $\mathcal{K}(t) \approx \mathcal{K}_2(t)$ in the stronger coupling regime. For simplicity, we consider noise coupling just strong enough that the memory kernel can be well approximated by $\mathcal{K}(t) \approx \mathcal{K}_2(t)+\mathcal{K}_4(t)$. We follow the prescription in Sec.~\ref{sec:specden2} to first extract the second-order memory kernel by using two sets of simulations under different $\omega_{s,i}$. In the lower panel of Fig.~(\ref{fig:toy-specden}), we plot the the theoretical and numerical values of the correlation functions $C_{zz}(t=15\delta t)$ at a specific time point versus the noise coupling strength. One clearly sees the increasing deviation of the numerically recovered correlation functions obtained under the second-order approximation as the noise coupling strength increases. However, the correct correlation function can still be recovered when we apply the method of Sec.~\ref{sec:specden2}.

\begin{figure}[htp]
\centering
\includegraphics[width=0.4\textwidth]{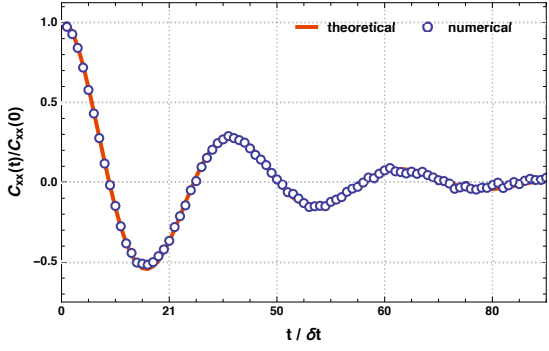}
\caption{TTM noise spectroscopy beyond the pure dephasing model. The numerically extracted correlation function (from the memory kernel) matches well with the theoretical result $C_{xx}(t)=\langle B^x(t)B^x(0) \rangle$. Model parameters: $H_s=0.02\sigma_z$,  $C_{xx}(0)=0.01$ and $dt=0.04$.
}
\label{fig:toy-specden2}
\end{figure}

Lastly, we present a model with population relaxation in order to demonstrate the wide applicability of TTM noise spectroscopy.
The Hamiltonian we consider is $H_s=\omega_s\sigma^z$ and $H_{sb}(t)= B_x(t)\sigma^x$.
As shown in Fig.~(\ref{fig:toy-specden2}), TTM noise spectroscopy also accurately reproduces the theoretical results for a non-pure-dephasing model, without requiring adjustments to the procedure.

%%%%%%%%%%%%%%%%%%%%%%%%%%%%%%%%%%%%%%%%%%%%%%%%%%%%%%%%%
\subsection{Two-Qubit Case}\label{sec:toy2}
%%%%%%%%%%%%%%%%%%%%%%%%%%%%%%%%%%%%%%%%%%%%%%%%%%%%%%%%%

We now turn to two-qubit spectroscopy, and consider two different models for comparison. In the first case, two coupled qubits are subject to independent noise, $H_s = \omega_{1}\sigma^z_1 + \omega_{2}\sigma^z_2+\omega_{12}\sigma^z_1\sigma^z_2$ and $\langle B_1^z(t)B_2^z(t^\prime)\rangle=0$. In the other case, the qubits are not directly coupled but are subject to correlated noise, $H_s = \omega_{1}\sigma^z_1 + \omega_{2}\sigma^z_2$ with $\langle B_1^z(t)B_2^z(t^\prime)\rangle\neq 0$. We shall refer to these as the independent (first) and correlated (second) models, respectively.

\begin{figure}[htp]
\centering
\includegraphics[width=0.38\textwidth]{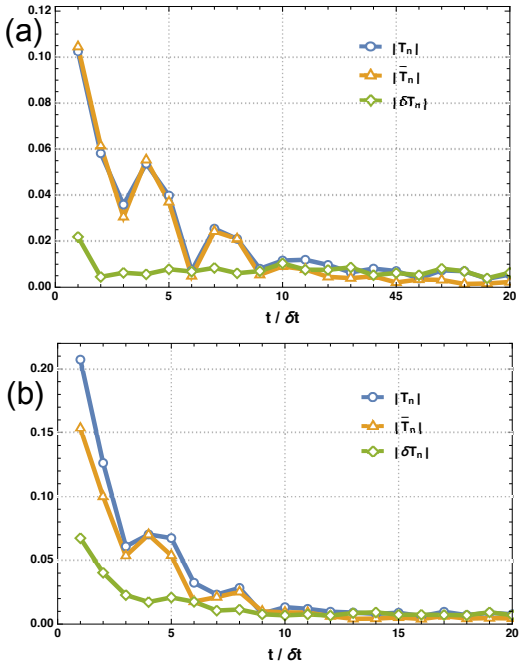}
\caption{TTM's Frobenious norm distributions $\vert T_n \vert$  for two different models explained in the text. We not only plot the norms for the full TTMs but also the norms for separable and correlated TTMs, respectively. (Top): The first model is $H_s=\sum_{i}\omega_{s,i}\sigma^z_i + \omega_{s,12}\sigma^z_1\sigma^z_2$, and the noise sources $B^z_1(t)$ and $B^z_2(t)$ are
localized and independent from each other. (Bottom): The second model has no explicit coupling term in $H_s$, i.e. $\omega_{s,12}=0$, with correlated noise sources.}
\label{fig:toy-ttm-fact}
\end{figure}

In Fig.~(\ref{fig:toy-ttm-fact}), we plot the norms of the full, separable, and correlated TTMs (i.e. $T_n$, $\bar{T}_n$, and $\delta T_n$, respectively) for the first model in the upper panel and the second model in the lower panel. For better presentation, we subtract an identity matrix from $T_1$ and $\bar{T}_1$ plotted in both panels of Fig.~(\ref{fig:toy-ttm-fact}). In the upper panel, only $\delta T_1$ has non-trivial norm. This result indicates that the correlated part of the TTM is ``almost" Markovian in nature when two qubits are subject to independent noise. One can further decompose $\delta T_1$ into $\delta \mathcal{L}_s$
and $\delta\mathcal{K}(t_1)$. This can be done because the definition $\delta T_1 = \delta \mathcal{L} \delta t + \delta \mathcal{K}(t_1) \delta t^2$ implies the two terms scale differently with respect to $\delta t$. Hence, we can construct two $\delta T_1$ (with slightly different time step size $\delta t$), and simple algebraic manipulations allow us to isolate the individual terms, $\delta \mathcal{L}$ or $\delta \mathcal{K}(t_1)$, making up $\delta T_1$. In this particular case, we confirm (cf. App.\ref{app:toy2}) that the $\delta \mathcal{L}_s$, reminiscent of the $\omega_{12}\sigma^z_1\sigma^z_2$ term, is responsible for entangling the two qubits and allows decoherence effects to be correlated even if the noise sources are spatially separated and independent. In the second case (with explicitly correlated noise), we plot the norms of different TTMs in the lower panel of Fig.~(\ref{fig:toy-ttm-fact}). Here, multiple $\delta T_n$ have sizable norm. A closer inspection on the magnitude of $\delta \mathcal{L}$ and $\delta \mathcal{K}(t_1)$ indicates that $\delta \mathcal{K}(t_1)$ is the main contributing factor for $\delta T_1$. Based on this straightforward analysis of TTM norm distribution, one can estimate the relative importance of different physical mechanisms contributing to collective decoherence.

\begin{figure}[htp]
\centering
\includegraphics[width=0.49\textwidth]{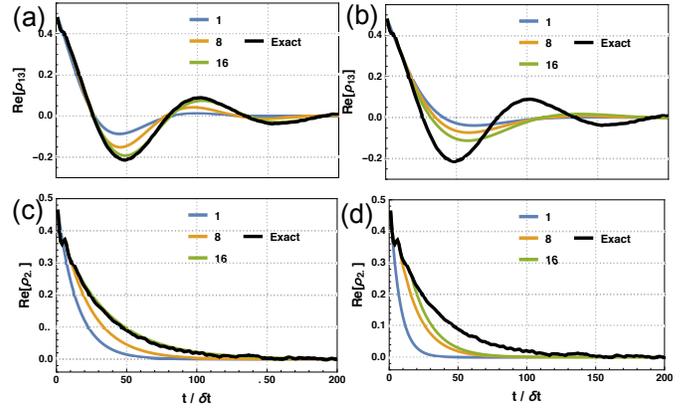}
\caption{TTM inference for two-qubit pure dephasing models. In all panels, we limit the number of dynamical maps used ($n=1$, $8$ and $16$) for the TTM construction and dynamical predictions on some chosen matrix elements of the two-qubit density matrix. Standard computational basis is used to label the matrix elements, i.e. $\ket{00} \rightarrow 1$ and $\ket{11}\rightarrow 4$ etc. (Left column): The temporal profile of coherence decay for the first model (upper panel) and the second model (lower panel) as predicted by the full TTMs.
(Right column): The temporal profile of coherence decay for the first model (upper panel) and the second model (lower panel) as predicted by the separable TTMs.
Details of the first model:$H_s=0.1\sigma^z_1+0.1\sigma^z_2+0.05\sigma^z_1\sigma^z_2$, $H_{sb}=B_1^z(t)\sigma^z_1+B_2^z(t)\sigma^z_2$, $C_{zz}(0)=\langle B_1^z(t) B_1^z(0)\rangle=\langle B_2^z(t) B_2^z(0)\rangle=1$ and $dt=0.2$. The initial state reads $\ket{\psi(0)}=(|00 \rangle+|10 \rangle)/\sqrt{2}$.
Details of the second model:
$H_s=0.1\sigma^z_1+0.1\sigma^z_2$, $H_{sb}=B_1^z(t)\sigma^z_1+B_2^z(t)\sigma^z_2$, $\langle B_1^z(t) B_1^z(0)\rangle=\langle B_2^z(t) B_2^z(0)\rangle=\langle B_1^z(t) B_2^z(0)\rangle=1$ and $dt=0.2$.
The initial state reads $\ket{\psi(0)}=(|01 \rangle+|10 \rangle)/\sqrt{2}$.}
\label{fig:toy-2qubit-predict}
\end{figure}

We next investigate the set of correlated TTMs $\delta T_n$.
Figure (\ref{fig:toy-2qubit-predict}) presents the dynamical evolution of an off-diagonal matrix elements of a two-qubit density matrix. Similar to the single-qubit case in Fig.~(\ref{fig:toy-ttmnorm-predict}), we compare the dynamics generated via TTM with the exactly simulated results.
Across the upper panel, we plot the dynamical prediction based on the full TTM (left column) and separable TTM (right column) against the exact results for the first model.
Across the lower panel, we compare the results of full TTM (left column) and separable TTM (right column) to the exact results for the second model.
In both cases, effects of collective decoherence are seen by the failure to correctly generate the dynamics using the separable TTM $\bar{T}_n$ alone. One may be deceived by  Fig.~(\ref{fig:toy-ttm-fact}) into thinking that $\delta T_n$ might not play important roles in both two models we consider here, as the norms for $\delta T_n$ are much smaller to those of $\bar{T}_n$. Yet, dropping these seemingly small terms results in inaccurate dynamical predictions as shown by the discrepancy between the left and right columns of Fig.~(\ref{fig:toy-2qubit-predict}).
This observation exemplifies the intricate nature of highly non-Markovian dynamics. The norm distribution of the full TTMs in both panels of Fig.~(\ref{fig:toy-ttm-fact}) suggests high non-Markovianity in both cases we investigate.

The model studies in this section illustrate the usefulness of using TTM spectroscopy
to distinguish between the underlying causes of collective decoherences in multi-qubit circuits. Improving the coherence time would require very different engineering efforts to either reduce a direct qubit-qubit interaction, $\sigma^z_1\sigma^z_2$, or to eliminate common noise sources.

%%%%%%%%%%%%%%%%%%%%%%%%%%%%%%%%%%%%%%%%%%%%%%%%%%%%%%%%%
\section{TTM on the IBM Quantum Experience}\label{sec:ibm}
%%%%%%%%%%%%%%%%%%%%%%%%%%%%%%%%%%%%%%%%%%%%%%%%%%%%%%%%%
We further test  TTM spectroscopy on IBM's 14-qubit device, ``IBM Q Melbourne".
In this device, a single-qubit pulse lasts 100 $ns$ with a buffer time of 10 $ns$ free evolution between pulses.  To construct the TTMs, we apply a sequence of `identity' gates to instruct the control system not to interfere with the qubit's free evolution in the presence of noise sources. Because of the aforementioned setup for this particular device, we always sample the QPTs at integer multiples of 110 $ns$, the combined time for a single-qubit rotation and an interim buffer.
%We also assume that the noise coupling is sufficiently weak for a high-quality hardware like IBM Q and take $\mathcal{K}(t) \approx \mathcal{K}_2(t)$ in the following
%analysis.

To obtain QPTs, we perform all possible n-qubit Pauli measurements $\sigma^{\alpha_1}_1\cdots\sigma^{\alpha_n}_n$
under $4^n$ initial conditions at every sampling time point. We denote $\alpha_i \in \{x,y,z,0\}$ with $\sigma^0 = I$, the identity operator.
We perform simple tests to demonstrate that TTM provides numerous insights to characterize noise properties even when only restrictive operations performed over the cloud are possible.  Additional details may be found in App.~\ref{app:ibm}.

%%%%%%%%%%%%%%%%%%%%%%%%%%%%%%%%%%%%%%%%%%%%%%%%%%%%%%%%%
\subsection{One-Qubit Case}
%%%%%%%%%%%%%%%%%%%%%%%%%%%%%%%%%%%%%%%%%%%%%%%%%%%%%%%%%

\begin{figure}[htp]
\centering
\includegraphics[width=0.35\textwidth]{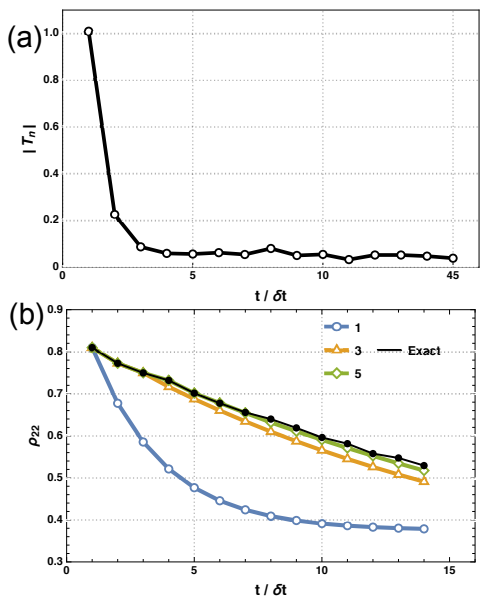}
\caption{TTM inference for a single-qubit experiment on IBM Q with $\delta t=2.2\mu s$. (Top): Frobenious norm distribution of the transfer tensor maps, $\vert T_n \vert$. (Bottom): The time-dependent population in the state $|1 \rangle$. TTM prediction based on $n=1$, $3$, and $5$ dynamical maps are presented.}
\label{fig:ibm-1-norm}
\end{figure}

We first convert the experimental QPTs collected from an IBM device into TTMs. The norm distribution is displayed in the upper panel of Fig.~(\ref{fig:ibm-1-norm}). We find the qubits are affected by non-Markovian noise, as there is more than one TTM with sizable norm in the plot.
%We attribute the non-vanishing tail associated with the higher-indexed TTMs to the state preparation and measurement errors of the device. For instance, we observe that a measurement immediately following an initialization $\rho(0)=\ket{1}\bra{1}$ does not give back exactly the state $\ket{1}$.
%To best fit the experimental outcomes with a consistent qubit model, we drop these erroneous measurement results, that should vanish, from the raw QPT data. After this minimal adjustments, the norm distribution of corrected TTMs decreases for higher-indexed maps as shown in Fig.~(\ref{fig:ibm-1-norm}).

\begin{figure}[htp]
\centering
\includegraphics[width=0.4\textwidth]{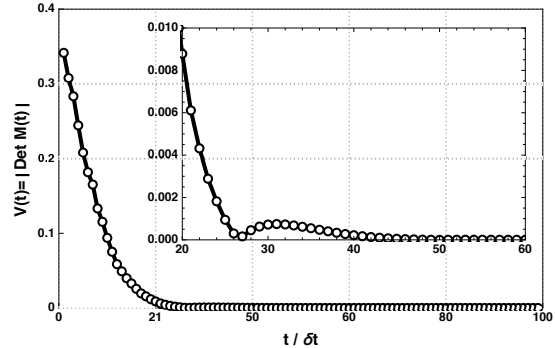}
\caption{The time-dependent Bloch volume $V(t)$ is plotted. The non-Markovianity is manifested by the temporary reversal of the monotonic decay trend of $V(t)$ in the inset.}
\label{fig:ibm-1-bloch}
\end{figure}

We next perform additional experiments to corroborate the usefulness of TTMs. First, we use $T_n$ to predict the dissipative quantum dynamics. As inferred from the upper panel of Fig.~(\ref{fig:ibm-1-norm}), a minimum of 3 TTMs is required to faithfully reproduce the dissipative effects arising from noise influences. In the lower panel, we do observe that significant errors accumulate quickly when fewer TTMs are used. Figure~(\ref{fig:ibm-1-norm}) suggest the memory kernel's time length lasts on the order of 1 $\mu$s. This is not particularly short in comparison to the gating time of 100 ns.

We further investigate other aspects of non-Markovianity by plotting the Bloch volume $V(t)=\text{det}\vert M(t) \vert$ in Fig.~(\ref{fig:ibm-1-bloch}). In this case, the temporary reversal of the Bloch volume contraction (inset of Fig.~(\ref{fig:ibm-1-bloch})), is observed far outside the time window for data acquisition. The first five data points in Fig.~(\ref{fig:ibm-1-bloch}) are experimentally measured results from the IBM device, the rest are numerically generated with the help of TTMs. Without relying on TTM predictions, it is not possible to obtain such a result given the circuit depth restriction on the IBM Q.
While we do observe a signature of non-Markovianity, the magnitude of the Bloch volume is already very small, as shown in the inset of Fig.~(\ref{fig:ibm-1-bloch}). Hence, the observed reversal of Bloch volume contraction does not imply a strong degree of non-Markovianity.

\begin{figure}[htp]
\centering
\includegraphics[width=0.45\textwidth]{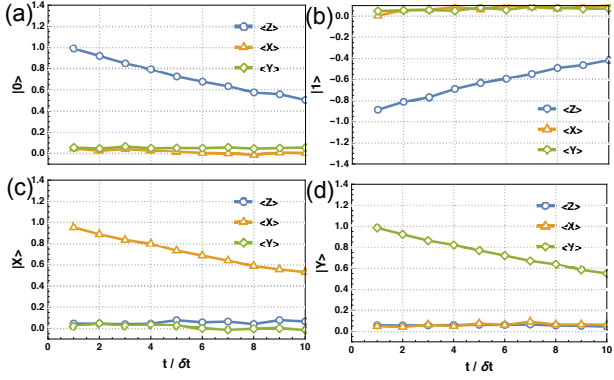}
\caption{TTM inference for single-qubit dynamical decoupling (DD) experiment on IBM Q with $\delta t=2.64\mu s$. Under the $XY4$ DD protocol, expectation values for all three Pauli operators are measured for four initial conditions.
(a): $\ket{\psi(0)}=\ket{0}$.
(b): $\ket{\psi(0)}=\ket{1}$.
(c): $\ket{\psi(0)}\propto \ket{0}+\ket{1}$.
(d): $\ket{\psi(0)}\propto \ket{0}+i\ket{1}$. }
\label{fig:ibm-xy}
\end{figure}

\begin{figure}[htp!]
\centering
\includegraphics[width=0.4\textwidth]{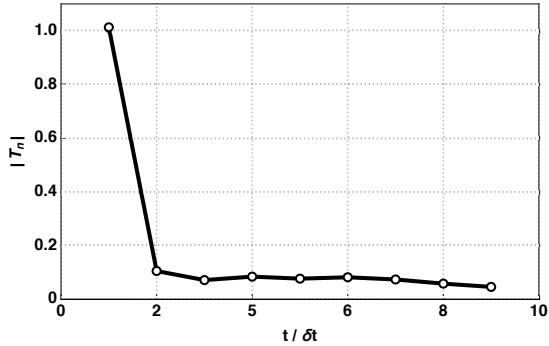}
\caption{Under the $XY4$ DD protocol, Frobenious norm distribution of the effective transfer tensor maps with $\delta t=2.64\mu$s.}
\label{fig:ibm-zz}
\end{figure}

Next, we draw attention to a recent work \cite{pokharel_demonstration_2018} by Pokharel et. al. to study the effects of DD on preserving qubit  coherence on both IBM and Rigetti devices.  They use the $XY4$ DD control scheme to prolong qubit coherence. We repeat the DD experiment as shown in in Fig.~(\ref{fig:ibm-xy}), and verify the prolonging of quantum coherence and the trapping of state populations. This change in the underlying dynamical process is reflected in the effective TTM norm distributed plotted in
(\ref{fig:ibm-zz}). We note that the effective noise under $XY4$ DD control scheme looks more Markovian with a sharper decay profile in the norm distribution in comparison to the original one (free evolution) given in Fig.~(\ref{fig:ibm-1-norm}). This is consistent with the observation given by Pokharel et. al.

%%%%%%%%%%%%%%%%%%%%%%%%%%%%%%%%%%%%%%%%%%%%%%%%%%%%%%%%%
\subsection{Two-Qubit Case}\label{sec:ibm2}
\begin{figure}  
\centering 
\includegraphics[width=0.35\textwidth]{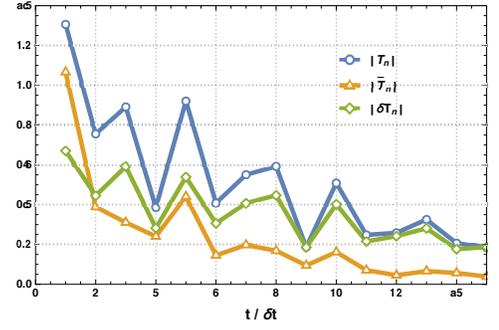}
\caption{TTM's Frobenious norm distributions for the two-qubit experiment on IBM Q with $\delta t=2.2\mu s$. 
For the two-qubit experiment, we not only plot the norms for the full TTM but also the norms for factorized and correlated TTMs, respectively.} 
\label{fig:ibm-2-norm}
\end{figure}

%%%%%%%%%%%%%%%%%%%%%%%%%%%%%%%%%%%%%%%%%%%%%%%%%%%%%%%%%
In two-qubit circuits we are primarily interested in detecting spatially correlated noise sources and their contributions decoherence, cf. the circuit layout in Fig.~(\ref{fig:ibm}). To this end, we plot the norms of full TTMs ($T_n$), separable TTMs ($\bar T_n$) and correlated TTMs ($\delta T_n$). Due to larger state preparation errors arising from the use of two-qubit gates, the TTM norm distribution stabilizes at a value much higher than 0 in Fig.~(\ref{fig:ibm-2-norm}) in comparison to the one-qubit results given in Fig.~(\ref{fig:ibm-1-norm}).  Nevertheless, the norm profile of first few TTMs suggest that correlated noise can play a significant role in the overall decoherence for the two-qubit circuit.
%The second TTM $\vert \delta T_2 \vert > \vert \bar T_2 \vert$ gives a crucial indication that the correlated noises could play an important role. 

\begin{figure}  
\centering 
\includegraphics[width=0.45\textwidth]{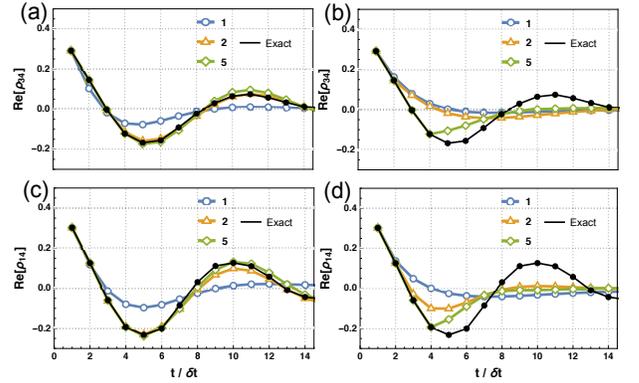}
\caption{TTM inference for the two-qubit experiments on IBM Q with $\delta t=2.2\mu s$.
In the upper / lower row, we limit the number of dynamical maps used ($n=1,2,4$ ) for the TTM construction and dynamical predictions on some chosen matrix elements of the two-qubit density matrix. Standard computational basis is used to label the matrix elements, i.e. $\ket{00} \rightarrow 1$ and $\ket{11}\rightarrow 4$ etc. (Left Column): Prediction of time-dependent quantum coherences based on the full TTMs for an unentangled initial state  $(|10 \rangle+|11 \rangle)/\sqrt{2}$ in the upper panel and an entangled initial state $(|00 \rangle+|11 \rangle)/\sqrt{2}$ in the lower panel. 
(Right Column): Corresponding prediction based on the separable TTM alone.} 
\label{fig:ibm-2-predict}
\end{figure}

To vindicate this hypothesis, we check  the dynamics inferred from the full and separable TTMs, respectively, against the experimental data shown in Fig.~(\ref{fig:ibm-2-predict}). The time evolution of a separable initial state, upper row in Fig.~(\ref{fig:ibm-2-predict}), is noticeably influenced by the collective decoherence because panel a (full TTM) and b (separable TTM) give inconsistent results. For an initially unentangled two-qubit system, there should not be any entanglement generation when being left alone as in this experiment. Yet, the strong disagreement between predictions by full and separable TTMs indicate that the two qubits must be subsequently entangled due to the action of $\delta T_n$.  
We also look at the concurrence for the two-qubit density matrix (directly obtained from state tomography) as an additional check to confirm the emergence of non-trivial entanglement after the state preparation stage at $t=0$. Unsurprisingly, the time evolution of an initially entangled state, lower row of Fig.~(\ref{fig:ibm-2-predict}), is also influenced by the collective decoherence as panel c (full TTM) and d (separable TTM) yield inconsistent outcomes. In both cases, $\delta T_n$ actually help to sustain quantum coherence. 

To identify the source of  collective decoherence in this circuit, we perform the analysis discussed in App.~(\ref{app:toy2}), i.e. we isolate $\delta \mathcal{L}_s$ and $\delta \mathcal{K}_1$ from  $\delta T_1$ and compare their norms. We find both terms to be non-negligible
with $\delta \mathcal{K}_1$ slightly dominant. Efforts to minimize the presence of collective noise ($\delta \mathcal{K}_1$) and suppression of cross-talk ($\delta \mathcal{L}_s$) are thus both needed to improve coherence on this particular device.

%%%%%%%%%%%%%%%%%%%%%%%%%%%%%%%%%%%%%%%
\section{Conclusion}
%%%%%%%%%%%%%%%%%%%%%%%%%%%%%%%%%%%%%%%
In this work, we proposed transfer tensor map based noise spectroscopy that enables physicists and engineers to more easily (1) characterize the non-Markovianity of a noise source, (2) determine the noise power spectrum, and (3) estimate spatial and temporal correlations of nonlocal noise a multi-qubit systems.  In addition, the TTM spectroscopy facilitates noise characterization beyond pure dephasing models. %and works for both quantum and classical noises.

The key idea is to tomographically reconstruct the well-established TNMQE with a memory kernel that encodes all information required to predict the effects of the noise. Furthermore, the second-order memory kernel is known to be proportional to the noise's autocorrelation functions. For high-quality quantum computing hardware, qubits
should be weakly coupled to noise. Therefore, one may often assume $\mathcal{K}(t)\approx \mathcal{K}_2(t)$ and deduce the correlation functions according to Eq.~(\ref{eq:ncorr}). The noise power spectrum can be subsequently generated. For non-weak-coupling cases, we propose a simple remedy to perform a series of experiments to extract the desired noise's correlation function in Sec.~\ref{sec:specden2}. Furthermore, we discuss how to unravel two-qubit TTMs to investigate the underlying nature of collective decoherence.

To illustrate different aspects of the proposed TTM spectroscopy, we applied it to simple theoretical models as well as IBM Q's hardware. For theoretical models,
we demonstrated an efficient scheme to assess the non-Markovianity of a quantum process, reconstructed the noise's correlation functions and clarified the source of collective decoherence for two-qubit circuits.  For the IBM Q hardware, the predictive power of TTM spectroscopy was unambiguously demonstrated. This implies that the TTMs encode realistic decoherence effects on the qubits. We show that the qubits undergo non-Markovian dissipation.
Finally, we observe that collective decoherence is due to the presence of spatially correlated noise sources as well as cross-talk between qubits in the ``IBM Q Melbourne" device.

%%%%%%%%%%%%%%%%%%%%%%%%%%%%%%%%%%%%%%%
\appendix
%%%%%%%%%%%%%%%%%%%%%%%%%%%%%%%%%%%%%%%%

%%%%%%%%%%%%%%%%%%%%%%%%%%%%%%%%%%%%%%%%
\section{Derivation of the Time-NonLocal Quantum Master Equation for Classical Noises}\label{app:proj}
%%%%%%%%%%%%%%%%%%%%%%%%%%%%%%%%%%%%%%%%
Given $H(t)$, the Hamiltonian in Eq.~(\ref{eq:H1}), we first consider the case of quantum noise. In this case, the projection $\mathcal{P}$ is defined via
\begin{eqnarray}\label{eq:qproj}
\mathcal{P}\Omega(t) = \text{Tr}_b(\Omega(t))\otimes\rho_b = \rho(t)\otimes\rho_b,
\end{eqnarray}
where $\Omega(t)$ is a joint system-bath quantum state and $\mathcal{P}+\mathcal{Q}=\mathcal{I}$ (the identity operator).
By introducing the projectors $\mathcal{P}$ and $\mathcal{Q}$ into the Liouville equation for the joint system, we obtain
\begin{eqnarray}
\frac{d}{dt}\mathcal{P}\Omega(t) & = & -i \mathcal{PLP}\Omega(t)-i\mathcal{PLQ}\Omega(t), \nonumber \\
\frac{d}{dt}\mathcal{Q}\Omega(t) & = & -i \mathcal{QLP}\Omega(t)-i\mathcal{QLQ}\Omega(t),
\end{eqnarray}
where $\mathcal{L}\Omega(t)=[H(t),\Omega(t)]$. We note that the equation of motion for $\mathcal{Q}\Omega(t)$ can be formally integrated.
By substituting the formal solution of $\mathcal{Q}\Omega(t)$ into the equation of motion for
$\mathcal{P}\Omega(t)$, one derives
\begin{eqnarray}
\frac{d}{dt}\mathcal{P}\Omega(t)&=&-i\mathcal{PLP}\Omega(t)-i\mathcal{PL}e^{-i\mathcal{QL}t}\mathcal{Q}\Omega(0) \nonumber \\
& & \, -\int^t_0 d\tau \mathcal{PLQ}e^{-i\mathcal{QL}(t-\tau)}\mathcal{QLP}\Omega(\tau).
\end{eqnarray}
Furthermore, by assuming $\mathcal{PLP} = 0$ (consistent with properties of Gaussian noise) and $\mathcal{Q}\Omega(0)=0$ (a factorized initial state of
the form $\rho(0)\otimes\rho_b$) the above equation reduces to the TNQME, Eq.~(\ref{eq:nz-mastereq}) with the memory kernel given in Eq.~(\ref{eq:memker}).

Next, we consider the case of classical noise. In this case, the system is not explicitly coupled to another quantum system acting as a source of noise.
Rather, the system is subjected to classical noise sources with the Hamiltonian $H(t)$ parameterized by Gaussian stochastic process $B^\alpha(t)$.
In this case, we define the projector $\mathcal{P}$ via
\begin{eqnarray}\label{eq:cproj}
\mathcal{P} \tilde\rho(t) = \langle \tilde\rho(t) \rangle = \rho(t),
\end{eqnarray}
where $\langle \cdots \rangle$ denotes an average over the Gaussian noise sources, while $\tilde\rho(t)$ is a stochastic instance of the system's density matrix.
Similarly, the complementary projector $\mathcal{Q}$ satisfy the relation $\mathcal{P}+\mathcal{Q}=\mathcal{I}$. By going through mathematical manipulation for the
quantum-noise case, one derives
\begin{eqnarray}
\frac{d}{dt}\mathcal{P}\tilde\rho(t)&=&-i\mathcal{PLP}\tilde\rho(t)-i\mathcal{PL}e^{-i\mathcal{QL}t}\mathcal{Q}\tilde\rho(0) \nonumber \\
& & \, -\int^t_0 d\tau \mathcal{PLQ}e^{-i\mathcal{QL}(t-\tau)}\mathcal{QLP}\tilde\rho(\tau),
\end{eqnarray}
where $\mathcal{L}\rho(t)=[H(t),\rho(t)]$.
Similarly, one has to make the same assumptions:  $\mathcal{PLP} = 0$ and $\mathcal{Q}\Omega(0)=0$; then one obtains formally the same TNQME, Eq.~(\ref{eq:nz-mastereq})
with a structurally identical memory kernel.  The subtle differences between quantum and classical cases are as follows. (1) $\mathcal{L}$ is a stochastic
superoperator acting on the system in the classical case, while  in the quantum setting it is a deterministic superoperator acting on the joint system and bath Hilbert space; and (2)
The definitions of the projection operator $\mathcal{P}$ differ in Eq.~(\ref{eq:qproj}) and (\ref{eq:cproj}).

Finally, to obtain the spectral density for the classical noise case, note that the second-order
memory kernel retains the functional form of Eq.~(\ref{eq:corr}) with the obvious condition that $C_{\alpha\alpha^\prime}(t)$ has to be strictly real-valued and  $C_{\alpha\alpha^\prime}(-t)=C_{\alpha'\alpha}(t)$ . In this case, the noise spectral density is obtained via the Wiener-Khinchin theorem \cite{clerk_devoret_rmp10} for stochastic processes,
\begin{eqnarray}
S_{\alpha\alpha^\prime}(\omega) & = & \int^\infty_{-\infty}dt e^{i\omega t} C_{\alpha\alpha^\prime}(t).
\end{eqnarray}
The spectral density obeys $S_{\alpha\alpha^\prime}(\omega)=S_{\alpha\alpha^\prime}(-\omega)$ for classical noise.

%%%%%%%%%%%%%%%%%%%%%%%%%%%%%%%%%
\section{Matrix Representations of Dynamical Maps}\label{app:MtxRep}
%%%%%%%%%%%%%%%%%%%%%%%%%%%%%%%%%
In Sec.~\ref{sec:specden2}, we implicitly assume the following matrix form \cite{shen_noh_prb17} for a dynamical map $\mathcal{T}$,
\begin{eqnarray}\label{eq:app-dynmap}
\mathcal{E}_{\mathbf{ij}}=\text{Tr}\left[\ket{i^\prime}\bra{i}\mathcal{T}(\ket{j}\bra{j^\prime})\right],
\end{eqnarray}
where $\mathbf{i}=(i,i^\prime)$, $i,i^\prime \in \{1,2,\cdots,d\}$ and $\ket{i}$ refers to one of the orthogonal set of quantum states for a $d$-level quantum system. The state transformation induced by the superoperator $\mathcal{E}$ (of dimension $d^2$ by $d^2$) can be operationally carried out as a matrix vector product when the density matrix $\rho$ is encoded as a vector of length $d^2$ in a Liouville space, i.e. $\rho^\prime_{ij}=\sum_{kl} \mathcal{E}_{ij,kl}\rho_{kl}$. The structural form of $\mathcal{E}$ brings many operational advantages. For instance, a successive transition (in the form of compositions of dynamical maps) can be done with simple matrix multiplications, $\mathcal{T}(\mathcal{T}(v)) \rightarrow \mathcal{E}^2\rho_v = \mathcal{E}\mathcal{E}\rho_v$.

However, the Choi matrix~\cite{caruso_rmp14,filippov_sergey_njp17}, an alternative representation of the dynamical process, possesses other advantages. In order for a Choi matrix to represent a valid dynamical map, the matrix has to satisfy the trace condition (sum up to dimension of the Hilbert space), positivity, and Hermiticity.  The Choi matrix $\mathcal{X}$ is defined via,
\begin{eqnarray}\label{eq:app-procmtx}
\mathcal{T}\left(\boldmath{\cdot}\right) = \sum_{\mathbf{kl}} \mathcal{X}_{\mathbf{kl}} A_\mathbf{k} \left(\boldmath{\cdot}\right) A_\mathbf{l}^\dag
\end{eqnarray}
where the operators $A_{\mathbf{k}}=\ket{k}\bra{k^\prime}$. The the
matrices defined in Eq.~(\ref{eq:app-dynmap}) and (\ref{eq:app-procmtx}) are related via
\begin{eqnarray}\label{eq:choi-dynmap}
\mathcal{E}_{(i,i^\prime),(j,j^\prime)} = \mathcal{X}_{(i,j),(i^\prime,j^\prime)}
\end{eqnarray}

For a bipartite quantum system (composed of two $d$-level subsystems), the common practice is to index the state as
$\ket{i} = \ket{i_1}\ket{i_2}$ with $i=(i_1-1)d+i_2$ and the Liouville-space basis element $A_\mathbf{i} =\ket{i}\bra{i^\prime}$ via $\mathbf{i}=(i-1)d^2+i^\prime$. For this study, we find it is useful
to re-index the Liouville-space basis element $A_{\mathbf{i}}$
by replacing the standard approach  $\mathbf{i} \rightarrow (i_1,i_2; i^\prime_1,i^\prime_2$  with $\mathbf{i} \rightarrow (i_1,i^\prime_1;i_2,i_2^\prime)$,
where $(x,y;a,b) \equiv ((x-1)d+y-1)d^2 + ((a-1)d + b)$ and $x$,$y$,$a$, $b \in \{1, \cdots d\}$.
In terms of re-indexed Liouville-space basis elements, we get a convenient form of separable quantum channels for a bipartite system,
\begin{eqnarray}
\mathcal{T}(\boldmath{\cdot}) =
\sum_{\mathbf{i}_1,\mathbf{i}_2,\mathbf{j}_1,\mathbf{j}_2} \mathcal{X}^1_{\mathbf{i}_1,\mathbf{j}_1}
 \mathcal{X}^2_{\mathbf{i}_2,\mathbf{j}_2}
A_{\mathbf{i}_1}A_{\mathbf{i}_2}(\boldmath{\cdot})A^\dag_{\mathbf{j}_1}A^\dag_{\mathbf{j}_2}.
\end{eqnarray}
If we define $\mathcal{X}_{\mathbf{i},\mathbf{j}} = \mathcal{X}^1_{\mathbf{i}_1,\mathbf{j}_1}\mathcal{X}^2_{\mathbf{i}_2,\mathbf{j}_2}$, i.e.
$\mathcal{X}=\mathcal{X}^1 \otimes \mathcal{X}^2$,
then we can recover the individual Choi matrices, such as $\mathcal{X}^1 = \text{Tr}_{2} \mathcal{X}/d$ and vice versa.

For a general bipartite quantum dynamical process, we can write any Choi matrix as $\mathcal{X} = \bar{\mathcal{X}}^1\otimes \bar{\mathcal{X}^2} + \delta \mathcal{X}$, where $\bar{\mathcal{X}}^1 = \text{Tr}_2 \mathcal{X} / d$. The correlated term $\delta \mathcal{X}$ vanishes exactly for separable quantum channels.
Based on Eq.~(\ref{eq:choi-dynmap}), we can re-arrange the matrix elements of various $\mathcal{X}$ matrices to obtain the corresponding
$\mathcal{E}$ matrices. In particular, the factorized dynamical map reads,
\begin{eqnarray}
\bar{\mathcal{E}}_{(i_1,i^\prime_1;i_2,i^\prime_2),(j_1,j^\prime_1; j_2,j^\prime_2)} &=& \bar{\chi}^1_{i_1,j_1;i^\prime_1,j^\prime_1} \bar{\chi}^2_{i_2,j_2;i^\prime_2,j^\prime_2} \nonumber \\
& = &  \bar{\mathcal{E}}^1_{i_1,i^\prime_1;j_1 j^\prime_1} \bar{\mathcal{E}}^2_{i_2,i^\prime_2;j_2 j^\prime_2}.
\end{eqnarray}

%%%%%%%%%%%%%%%%%%%%%%%%%%%%%%%%%
\section{TTM Spectroscopy on Two-qubit Theoretical Models} \label{app:toy2}
%%%%%%%%%%%%%%%%%%%%%%%%%%%%%%%%%%
To isolate the $\delta \mathcal{L}$ and $\delta \mathcal{K}(t_1)$ terms, we construct two $T_1$ with different time step sizes $\delta t$ and $2\delta t$, respectively.  This gives
\begin{eqnarray}
\delta T_1&=&\delta \mathcal{L}\delta t +
\delta \mathcal{K}(t_1)\delta t^2       \nonumber \\
\delta T_1'&=&\delta \mathcal{L}(2\delta t) +
\delta \mathcal{K}'(t_1)(2\delta t)^2.
\end{eqnarray}
Since  $\delta t$ is small ($\delta t=0.2 <<$ the correlation length of noise $\approx 20$), we take $\mathcal{K}'(t_1)\approx \mathcal{K}(t_1) $. This approximation leads to
\begin{eqnarray}\label{eq:isolationLK}
\delta \mathcal{L}\delta t &=&(4\delta T_1-\delta T_1')/2,    \nonumber
\\
\delta \mathcal{K}(t_1)\delta t^2&=&-(2\delta T_1-\delta T_1')/2 . 
\end{eqnarray}

By substituting the transfer tensors $\delta T_1$ and $\delta T_1^\prime$ obtained for the first model (coupled qubits subjected to independent noises) in Sec.~\ref{sec:toy2} into Eqs.~(\ref{eq:isolationLK}), 
we find 
\begin{eqnarray}
\delta \mathcal{L}\delta t[2,2] &\approx &-0.0189241 i ,\\ \nonumber
\delta \mathcal{L}\delta t[3,3] &\approx &-0.0194065 i, \\ \nonumber
\delta \mathcal{L}\delta t[5,5] &\approx &0.0189241 i, \\ \nonumber
\delta \mathcal{L}\delta t[8,8] &\approx &0.0194065 i, \\ \nonumber
\delta \mathcal{L}\delta t[9,9] &\approx &0.0194065 i, \\   \nonumber
\delta \mathcal{L}\delta t[12,12] &\approx &0.0189241 i, \\ \nonumber
\delta \mathcal{L}\delta t[14,14] &\approx &-0.0194065 i, \\ \nonumber
\delta \mathcal{L}\delta t[15,15] &\approx &-0.0189241 i,  \\  \nonumber
\delta \mathcal{K}(t_1)\delta t^2&\approx & \mathbf{0} , \nonumber
\end{eqnarray}
which agrees well with the theoretical results of $[-i \omega_{s,12}\sigma_1^z \sigma_2^z,\rho]\delta t$ as $\delta \mathcal{L}\delta t[2,2]=\delta \mathcal{L}\delta t[3,3]=-\delta \mathcal{L}\delta t[5,5]=-\delta \mathcal{L}\delta t[8,8]=-\delta \mathcal{L}\delta t[9,9]=-\delta \mathcal{L}\delta t[12,12]=\delta \mathcal{L}\delta t[14,14]=\delta \mathcal{L}\delta t[15,15]=-i2\omega_{s,12}\delta t=-0.02i$

Similarly, if we substitute $\delta T_1$ and $T^\prime_1$ for the second model (independent qubits subjected to correlated noises) into Eqs.~(\ref{eq:isolationLK}), we find 
\begin{eqnarray}
\delta \mathcal{L}\delta t&\approx &\mathbf{0},  \\ \nonumber
\delta \mathcal{K}(t_1)\delta t^2[6,6]&\approx &-0.0594839,  \\ \nonumber
\delta \mathcal{K}(t_1)\delta t^2[7,7]&\approx &0.0637997,  \\ \nonumber
\delta \mathcal{K}(t_1)\delta t^2[10,10]&\approx & 0.0637997,  \\ \nonumber
\delta \mathcal{K}(t_1)\delta t^2[11,11]&\approx &-0.0594839, 
\end{eqnarray}
which agree well with the theoretical results for a correlated noise kernel $\mathcal{K}^{ZZ}(t_1\approx0)\delta t^2$ as $-\<B_1^z(0)B_2^z(0)\>\delta t^2[6,6]=\<B_1^z(0)B_2^z(0)\>\delta t^2[7,7]=\<B_1^z(0)B_2^z(0)\>\delta t^2[10,10]=-\<B_1^z(0)B_2^z(0)\>\delta t^2[11,11]=1*0.2^2=0.04$.

As commented at the end of Sec.~\ref{sec:specden2}, one can very accurately unravel the underlying causes for a collective decoherence in a multi-qubit circuit by following the analysis on the correlated transfer tensor maps outlined in this appendix.

%%%%%%%%%%%%%%%%%%%%%%%%%%%%%%%%%
\section{Quantum Process Tomography on the IBM Quantum Experience}\label{app:ibm}
%%%%%%%%%%%%%%%%%%%%%%%%%%%%%%%%%%
%In order to perform Eq.~(\ref{eq:ttm}) to get transfer tensor matrices to promote TTM process, we need the trajectories information, i.e. constructing dynamical map, by doing quantum process tomography.
%%%%%%%%%%%%%%%%%%%%%%%%%%%%%%%%%%%%%%%%%%%这个公式前面已经编号，删掉
%\begin{eqnarray}%\label{eq:ttm}
%T_n \equiv \mathcal{E}_n - \sum_{m=1}^{n-1} T_{n-m}\mathcal{E}_m,
%\end{eqnarray}

%\begin{eqnarray}\label{eq:dynamic map}
%\rho(t)=\mathcal{E}_t \rho(0).
%\end{eqnarray}

\begin{figure} 
\centering 
\includegraphics[width=0.45\textwidth]{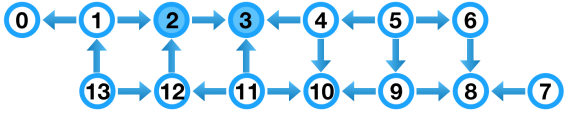}
\caption{Topology of ``IBM Q Melbourne". Arrows denote control directions of controlled-U gate. The shaded nodes 2 and 3 are the neighboring qubits used in the two-qubit experiments reported in Sec.~\ref{sec:ibm2}.} 
\label{fig:ibm}
\end{figure}

The experiments reported in Sec.~\ref{sec:ibm} were performed on a 14-qubit chip. The connectivity between qubits is illustrated in Fig.~(\ref{fig:ibm}). Below, we briefly describe the set of quantum states used to obtain dynamical maps via QPT. 

%Experiments in main test were performed on 4/1/19, during which for qubit[2] "u2" gate error is 0.003774135833008896 and  "u3" gate error is 0.00754827166601779, for qubit[3] "u2" gate error is 0.0011156577216821506 and  "u3" gate error is 0.002231315443364301.  Dephasing times (T2) for the qubits were between 40 − 100μs.

Dynamical maps are a linear, completely positive, trace preserving transformations. Quantum process tomography (QPT) is the protocol to determine the dynamical map evolving a general state from one point in time to another. This procedure requires one to perform a series of experiments involving various combinations of initial states and measurements. If the system under consideration has dimension d, we may choose $d^2$ pure sates $|\psi_1\rangle,...,|\psi_{d^2}\rangle$  for the QPT reconstruction. For an one-qubit system, we prepare the following four states as the basis for
experiments run on the IBM Q:
\newline
\begin{eqnarray}\label{eq:qpt1}
\vert \psi_0 \rangle &= & \vert 0 \rangle, \, \,
\vert \psi_x \rangle =  (\vert 0 \rangle + \vert 1 \rangle)/\sqrt{2}, \nonumber \\
\vert \psi_1 \rangle & = & \vert 1 \rangle, \,\,
\vert \psi_y \rangle = \vert 0 \rangle +i \vert 1 \rangle)/\sqrt{2}.
\end{eqnarray}
For a two-qubit system, we choose the following sixteen states as the basis:
\begin{widetext}
\begin{eqnarray}\label{eq:qpt2}
\vert \psi_{00} \rangle &=& \vert 00 \rangle, \qquad
\vert \psi_{01} \rangle = \vert 01 \rangle, \qquad
\vert \psi_{10} \rangle = \vert 10 \rangle, \qquad
\vert \psi_{11} \rangle = \vert 11 \rangle, \\
\vert \psi_{0X} \rangle &=& \frac{1}{\sqrt{2}} (\vert 00 \rangle+\vert 01 \rangle), \,\,
\vert \psi_{0Y} \rangle = \frac{1}{\sqrt{2}}(\vert 00 \rangle+i \vert 01 \rangle), \,\,
\vert \psi_{1X} \rangle = \frac{1}{\sqrt{2}}(\vert 10 \rangle+\vert 11 \rangle), \,\,
\vert \psi_{1Y} \rangle = \frac{1}{\sqrt{2}}(\vert 10 \rangle+i \vert 11 \rangle), \nonumber \\
\vert \psi_{X0} \rangle &=&  \frac{1}{\sqrt{2}} (\vert 00 \rangle+\vert 10 \rangle), \,\,
\vert \psi_{Y0} \rangle = \frac{1}{\sqrt{2}}(\vert 00 \rangle+i \vert 10 \rangle), \,\,
\vert \psi_{X1} \rangle = \frac{1}{\sqrt{2}}(\vert 11 \rangle+\vert 01 \rangle), \,\,
\vert \psi_{Y1} \rangle = \frac{1}{\sqrt{2}}(\vert 01 \rangle+i \vert 11 \rangle), \nonumber \\
\vert  \Phi \rangle  &=& \frac{1}{\sqrt{2}}(\vert 00 \rangle + \vert 11 \rangle), \,\,
\vert  \Psi \rangle = \frac{1}{\sqrt{2}}(\vert 01 \rangle +  \vert 10 \rangle), \,\,
\vert  \Phi^* \rangle = \frac{1}{\sqrt{2}}(\vert 00 \rangle + i\vert 11 \rangle), \,\,
\vert  \Psi^* \rangle = \frac{1}{\sqrt{2}}(\vert 01 \rangle +  i\vert 10 \rangle). \nonumber
\end{eqnarray}
\end{widetext}

We note the dynamical maps introduced in Sec.~\ref{sec:ttm} are matrices with respect to the
standard basis $\{\ket{n}\bra{m}\}$ where $n, m = 1, \dots d$ for $d=2$ (one-qubit case) or
$d=4$ (two-qubit case).  Because quantum dynamical processes are linear maps, it is straightforward to convert the experimentally reconstructed QPTs into dynamical maps in the standard basis.

%\%bibliographystyle{apsrev4-1}
%\bibliography{ref}

\begin{thebibliography}{65}%
\makeatletter
\providecommand \@ifxundefined [1]{%
 \@ifx{#1\undefined}
}%
\providecommand \@ifnum [1]{%
 \ifnum #1\expandafter \@firstoftwo
 \else \expandafter \@secondoftwo
 \fi
}%
\providecommand \@ifx [1]{%
 \ifx #1\expandafter \@firstoftwo
 \else \expandafter \@secondoftwo
 \fi
}%
\providecommand \natexlab [1]{#1}%
\providecommand \enquote  [1]{``#1''}%
\providecommand \bibnamefont  [1]{#1}%
\providecommand \bibfnamefont [1]{#1}%
\providecommand \citenamefont [1]{#1}%
\providecommand \href@noop [0]{\@secondoftwo}%
\providecommand \href [0]{\begingroup \@sanitize@url \@href}%
\providecommand \@href[1]{\@@startlink{#1}\@@href}%
\providecommand \@@href[1]{\endgroup#1\@@endlink}%
\providecommand \@sanitize@url [0]{\catcode `\\12\catcode `\$12\catcode
  `\&12\catcode `\#12\catcode `\^12\catcode `\_12\catcode `\%12\relax}%
\providecommand \@@startlink[1]{}%
\providecommand \@@endlink[0]{}%
\providecommand \url  [0]{\begingroup\@sanitize@url \@url }%
\providecommand \@url [1]{\endgroup\@href {#1}{\urlprefix }}%
\providecommand \urlprefix  [0]{URL }%
\providecommand \Eprint [0]{\href }%
\providecommand \doibase [0]{http://dx.doi.org/}%
\providecommand \selectlanguage [0]{\@gobble}%
\providecommand \bibinfo  [0]{\@secondoftwo}%
\providecommand \bibfield  [0]{\@secondoftwo}%
\providecommand \translation [1]{[#1]}%
\providecommand \BibitemOpen [0]{}%
\providecommand \bibitemStop [0]{}%
\providecommand \bibitemNoStop [0]{.\EOS\space}%
\providecommand \EOS [0]{\spacefactor3000\relax}%
\providecommand \BibitemShut  [1]{\csname bibitem#1\endcsname}%
\let\auto@bib@innerbib\@empty
%</preamble>
\bibitem [{\citenamefont {Preskill}(2018)}]{preskill_qm_18}%
  \BibitemOpen
  \bibfield  {author} {\bibinfo {author} {\bibfnamefont {J.}~\bibnamefont
  {Preskill}},\ }\href@noop {} {\bibfield  {journal} {\bibinfo  {journal}
  {Quantum}\ }\textbf {\bibinfo {volume} {2}},\ \bibinfo {pages} {79} (\bibinfo
  {year} {2018})}\BibitemShut {NoStop}%
\bibitem [{\citenamefont {Breuer}\ and\ \citenamefont
  {Petruccione}(2007)}]{breuer_theory_2007}%
  \BibitemOpen
  \bibfield  {author} {\bibinfo {author} {\bibfnamefont {H.-P.}\ \bibnamefont
  {Breuer}}\ and\ \bibinfo {author} {\bibfnamefont {F.}~\bibnamefont
  {Petruccione}},\ }\href@noop {} {\emph {\bibinfo {title} {The Theory of Open
  Quantum Systems}}}\ (\bibinfo  {publisher} {Oxford University Press},\
  \bibinfo {year} {2007})\BibitemShut {NoStop}%
\bibitem [{\citenamefont {de~Vega}\ and\ \citenamefont
  {Alonso}(2015)}]{de_vega_dynamics_2015}%
  \BibitemOpen
  \bibfield  {author} {\bibinfo {author} {\bibfnamefont {I.}~\bibnamefont
  {de~Vega}}\ and\ \bibinfo {author} {\bibfnamefont {D.}~\bibnamefont
  {Alonso}},\ }\href@noop {} {\bibfield  {journal} {\bibinfo  {journal} {Rev.
  Mod. Phys.}\ }\textbf {\bibinfo {volume} {89}},\ \bibinfo {pages} {015001}
  (\bibinfo {year} {2015})}\BibitemShut {NoStop}%
\bibitem [{\citenamefont {Breuer}\ \emph {et~al.}(2016)\citenamefont {Breuer},
  \citenamefont {Laine}, \citenamefont {Piilo},\ and\ \citenamefont
  {Vacchini}}]{breuer_colloquium:_2016}%
  \BibitemOpen
  \bibfield  {author} {\bibinfo {author} {\bibfnamefont {H.-P.}\ \bibnamefont
  {Breuer}}, \bibinfo {author} {\bibfnamefont {E.-M.}\ \bibnamefont {Laine}},
  \bibinfo {author} {\bibfnamefont {J.}~\bibnamefont {Piilo}}, \ and\ \bibinfo
  {author} {\bibfnamefont {B.}~\bibnamefont {Vacchini}},\ }\href@noop {}
  {\bibfield  {journal} {\bibinfo  {journal} {Rev. Mod. Phys}\ }\textbf
  {\bibinfo {volume} {88}} (\bibinfo {year} {2016})}\BibitemShut {NoStop}%
\bibitem [{\citenamefont {Cerrillo}\ and\ \citenamefont
  {Cao}(2014)}]{PhysRevLett.112.110401}%
  \BibitemOpen
  \bibfield  {author} {\bibinfo {author} {\bibfnamefont {J.}~\bibnamefont
  {Cerrillo}}\ and\ \bibinfo {author} {\bibfnamefont {J.}~\bibnamefont {Cao}},\
  }\href@noop {} {\bibfield  {journal} {\bibinfo  {journal} {Phys. Rev. Lett.}\
  }\textbf {\bibinfo {volume} {112}},\ \bibinfo {pages} {110401} (\bibinfo
  {year} {2014})}\BibitemShut {NoStop}%
\bibitem [{\citenamefont {Chuang}\ and\ \citenamefont
  {Nielsen}(1997)}]{chuang_prescription_1997}%
  \BibitemOpen
  \bibfield  {author} {\bibinfo {author} {\bibfnamefont {I.~L.}\ \bibnamefont
  {Chuang}}\ and\ \bibinfo {author} {\bibfnamefont {M.~A.}\ \bibnamefont
  {Nielsen}},\ }\href@noop {} {\bibfield  {journal} {\bibinfo  {journal} {J.
  Mod. Opt.}\ }\textbf {\bibinfo {volume} {44}},\ \bibinfo {pages} {2455}
  (\bibinfo {year} {1997})}\BibitemShut {NoStop}%
\bibitem [{\citenamefont {Poyatos}\ \emph {et~al.}(1997)\citenamefont
  {Poyatos}, \citenamefont {Cirac},\ and\ \citenamefont
  {Zoller}}]{poyatos_complete_1997}%
  \BibitemOpen
  \bibfield  {author} {\bibinfo {author} {\bibfnamefont {J.~F.}\ \bibnamefont
  {Poyatos}}, \bibinfo {author} {\bibfnamefont {J.~I.}\ \bibnamefont {Cirac}},
  \ and\ \bibinfo {author} {\bibfnamefont {P.}~\bibnamefont {Zoller}},\
  }\href@noop {} {\bibfield  {journal} {\bibinfo  {journal} {Phys. Rev. Lett.}\
  }\textbf {\bibinfo {volume} {78}},\ \bibinfo {pages} {390} (\bibinfo {year}
  {1997})}\BibitemShut {NoStop}%
\bibitem [{\citenamefont {Nielsen}\ and\ \citenamefont
  {Chuang}(2000)}]{nielsen_chuang_book00}%
  \BibitemOpen
  \bibfield  {author} {\bibinfo {author} {\bibfnamefont {M.~A.}\ \bibnamefont
  {Nielsen}}\ and\ \bibinfo {author} {\bibfnamefont {I.~L.}\ \bibnamefont
  {Chuang}},\ }\href@noop {} {\emph {\bibinfo {title} {Quantum computation and
  quantum information}}}\ (\bibinfo  {publisher} {Cambridge University Press},\
  \bibinfo {address} {Cambridge, England},\ \bibinfo {year} {2000})\BibitemShut
  {NoStop}%
\bibitem [{\citenamefont {Bialczak}\ \emph {et~al.}(2010)\citenamefont
  {Bialczak}, \citenamefont {Ansmann}, \citenamefont {Hofheinz}, \citenamefont
  {Lucero}, \citenamefont {Neeley}, \citenamefont {O‚ÄôConnell}, \citenamefont
  {Sank}, \citenamefont {Wang}, \citenamefont {Wenner}, \citenamefont
  {Steffen}, \citenamefont {Cleland},\ and\ \citenamefont
  {Martinis}}]{bialczak_quantum_2010}%
  \BibitemOpen
  \bibfield  {author} {\bibinfo {author} {\bibfnamefont {R.~C.}\ \bibnamefont
  {Bialczak}}, \bibinfo {author} {\bibfnamefont {M.}~\bibnamefont {Ansmann}},
  \bibinfo {author} {\bibfnamefont {M.}~\bibnamefont {Hofheinz}}, \bibinfo
  {author} {\bibfnamefont {E.}~\bibnamefont {Lucero}}, \bibinfo {author}
  {\bibfnamefont {M.}~\bibnamefont {Neeley}}, \bibinfo {author} {\bibfnamefont
  {A.~D.}\ \bibnamefont {O‚ÄôConnell}}, \bibinfo {author} {\bibfnamefont
  {D.}~\bibnamefont {Sank}}, \bibinfo {author} {\bibfnamefont {H.}~\bibnamefont
  {Wang}}, \bibinfo {author} {\bibfnamefont {J.}~\bibnamefont {Wenner}},
  \bibinfo {author} {\bibfnamefont {M.}~\bibnamefont {Steffen}}, \bibinfo
  {author} {\bibfnamefont {A.~N.}\ \bibnamefont {Cleland}}, \ and\ \bibinfo
  {author} {\bibfnamefont {J.~M.}\ \bibnamefont {Martinis}},\ }\href {\doibase
  10.1038/nphys1639} {\bibfield  {journal} {\bibinfo  {journal} {Nat. Phys.}\
  }\textbf {\bibinfo {volume} {6}},\ \bibinfo {pages} {409} (\bibinfo {year}
  {2010})}\BibitemShut {NoStop}%
\bibitem [{\citenamefont {Yamamoto}\ \emph {et~al.}(2010)\citenamefont
  {Yamamoto}, \citenamefont {Neeley}, \citenamefont {Lucero}, \citenamefont
  {Bialczak}, \citenamefont {Kelly}, \citenamefont {Lenander}, \citenamefont
  {Mariantoni}, \citenamefont {O'Connell}, \citenamefont {Sank}, \citenamefont
  {Wang}, \citenamefont {Weides}, \citenamefont {Wenner}, \citenamefont {Yin},
  \citenamefont {Cleland},\ and\ \citenamefont
  {Martinis}}]{yamamoto_quantum_2010}%
  \BibitemOpen
  \bibfield  {author} {\bibinfo {author} {\bibfnamefont {T.}~\bibnamefont
  {Yamamoto}}, \bibinfo {author} {\bibfnamefont {M.}~\bibnamefont {Neeley}},
  \bibinfo {author} {\bibfnamefont {E.}~\bibnamefont {Lucero}}, \bibinfo
  {author} {\bibfnamefont {R.~C.}\ \bibnamefont {Bialczak}}, \bibinfo {author}
  {\bibfnamefont {J.}~\bibnamefont {Kelly}}, \bibinfo {author} {\bibfnamefont
  {M.}~\bibnamefont {Lenander}}, \bibinfo {author} {\bibfnamefont
  {M.}~\bibnamefont {Mariantoni}}, \bibinfo {author} {\bibfnamefont {A.~D.}\
  \bibnamefont {O'Connell}}, \bibinfo {author} {\bibfnamefont {D.}~\bibnamefont
  {Sank}}, \bibinfo {author} {\bibfnamefont {H.}~\bibnamefont {Wang}}, \bibinfo
  {author} {\bibfnamefont {M.}~\bibnamefont {Weides}}, \bibinfo {author}
  {\bibfnamefont {J.}~\bibnamefont {Wenner}}, \bibinfo {author} {\bibfnamefont
  {Y.}~\bibnamefont {Yin}}, \bibinfo {author} {\bibfnamefont {A.~N.}\
  \bibnamefont {Cleland}}, \ and\ \bibinfo {author} {\bibfnamefont {J.~M.}\
  \bibnamefont {Martinis}},\ }\href@noop {} {\bibfield  {journal} {\bibinfo
  {journal} {Phys. Rev. B}\ }\textbf {\bibinfo {volume} {82}},\ \bibinfo
  {pages} {184515} (\bibinfo {year} {2010})}\BibitemShut {NoStop}%
\bibitem [{\citenamefont {Rodionov}\ \emph {et~al.}(2014)\citenamefont
  {Rodionov}, \citenamefont {Veitia}, \citenamefont {Barends}, \citenamefont
  {Kelly}, \citenamefont {Sank}, \citenamefont {Wenner}, \citenamefont
  {Martinis}, \citenamefont {Kosut},\ and\ \citenamefont
  {Korotkov}}]{rodionov_dnrey_prb2014}%
  \BibitemOpen
  \bibfield  {author} {\bibinfo {author} {\bibfnamefont {A.~V.}\ \bibnamefont
  {Rodionov}}, \bibinfo {author} {\bibfnamefont {A.}~\bibnamefont {Veitia}},
  \bibinfo {author} {\bibfnamefont {R.}~\bibnamefont {Barends}}, \bibinfo
  {author} {\bibfnamefont {J.}~\bibnamefont {Kelly}}, \bibinfo {author}
  {\bibfnamefont {D.}~\bibnamefont {Sank}}, \bibinfo {author} {\bibfnamefont
  {J.}~\bibnamefont {Wenner}}, \bibinfo {author} {\bibfnamefont {J.~M.}\
  \bibnamefont {Martinis}}, \bibinfo {author} {\bibfnamefont {R.~L.}\
  \bibnamefont {Kosut}}, \ and\ \bibinfo {author} {\bibfnamefont {A.~N.}\
  \bibnamefont {Korotkov}},\ }\href {\doibase 10.1103/PhysRevB.90.144504}
  {\bibfield  {journal} {\bibinfo  {journal} {Phys. Rev. B}\ }\textbf {\bibinfo
  {volume} {90}},\ \bibinfo {pages} {144504} (\bibinfo {year}
  {2014})}\BibitemShut {NoStop}%
\bibitem [{\citenamefont {Yuen-Zhou}\ \emph {et~al.}(2011)\citenamefont
  {Yuen-Zhou}, \citenamefont {Krich}, \citenamefont {Mohseni},\ and\
  \citenamefont {Aspuru-Guzik}}]{joelpnas11}%
  \BibitemOpen
  \bibfield  {author} {\bibinfo {author} {\bibfnamefont {J.}~\bibnamefont
  {Yuen-Zhou}}, \bibinfo {author} {\bibfnamefont {J.~J.}\ \bibnamefont
  {Krich}}, \bibinfo {author} {\bibfnamefont {M.}~\bibnamefont {Mohseni}}, \
  and\ \bibinfo {author} {\bibfnamefont {A.}~\bibnamefont {Aspuru-Guzik}},\
  }\href@noop {} {\bibfield  {journal} {\bibinfo  {journal} {Proc. Natl. Acad.
  Sci.}\ }\textbf {\bibinfo {volume} {108}},\ \bibinfo {pages} {17615}
  (\bibinfo {year} {2011})}\BibitemShut {NoStop}%
\bibitem [{\citenamefont {Howard}\ \emph {et~al.}(2006)\citenamefont {Howard},
  \citenamefont {Twamley}, \citenamefont {Wittmann}, \citenamefont {Gaebel},
  \citenamefont {Jelezko},\ and\ \citenamefont
  {Wrachtrup}}]{howard_quantum_2006}%
  \BibitemOpen
  \bibfield  {author} {\bibinfo {author} {\bibfnamefont {M.}~\bibnamefont
  {Howard}}, \bibinfo {author} {\bibfnamefont {J.}~\bibnamefont {Twamley}},
  \bibinfo {author} {\bibfnamefont {C.}~\bibnamefont {Wittmann}}, \bibinfo
  {author} {\bibfnamefont {T.}~\bibnamefont {Gaebel}}, \bibinfo {author}
  {\bibfnamefont {F.}~\bibnamefont {Jelezko}}, \ and\ \bibinfo {author}
  {\bibfnamefont {J.}~\bibnamefont {Wrachtrup}},\ }\href@noop {} {\bibfield
  {journal} {\bibinfo  {journal} {New J. Phys.}\ }\textbf {\bibinfo {volume}
  {8}},\ \bibinfo {pages} {33} (\bibinfo {year} {2006})}\BibitemShut {NoStop}%
\bibitem [{\citenamefont {Kofman}\ and\ \citenamefont
  {Korotkov}(2009)}]{kofman_two-qubit_2009}%
  \BibitemOpen
  \bibfield  {author} {\bibinfo {author} {\bibfnamefont {A.~G.}\ \bibnamefont
  {Kofman}}\ and\ \bibinfo {author} {\bibfnamefont {A.~N.}\ \bibnamefont
  {Korotkov}},\ }\href@noop {} {\bibfield  {journal} {\bibinfo  {journal}
  {Phys. Rev. A}\ }\textbf {\bibinfo {volume} {80}},\ \bibinfo {pages} {042103}
  (\bibinfo {year} {2009})}\BibitemShut {NoStop}%
\bibitem [{\citenamefont {Buser}\ \emph {et~al.}(2017)\citenamefont {Buser},
  \citenamefont {Cerrillo}, \citenamefont {Schaller},\ and\ \citenamefont
  {Cao}}]{buser_initial_2017}%
  \BibitemOpen
  \bibfield  {author} {\bibinfo {author} {\bibfnamefont {M.}~\bibnamefont
  {Buser}}, \bibinfo {author} {\bibfnamefont {J.}~\bibnamefont {Cerrillo}},
  \bibinfo {author} {\bibfnamefont {G.}~\bibnamefont {Schaller}}, \ and\
  \bibinfo {author} {\bibfnamefont {J.}~\bibnamefont {Cao}},\ }\href@noop {}
  {\bibfield  {journal} {\bibinfo  {journal} {Phys. Rev. A}\ }\textbf {\bibinfo
  {volume} {96}},\ \bibinfo {pages} {062122} (\bibinfo {year}
  {2017})}\BibitemShut {NoStop}%
\bibitem [{\citenamefont {Kananenka}\ \emph {et~al.}(2016)\citenamefont
  {Kananenka}, \citenamefont {Hsieh}, \citenamefont {Cao},\ and\ \citenamefont
  {Geva}}]{kananenka_accurate_2016}%
  \BibitemOpen
  \bibfield  {author} {\bibinfo {author} {\bibfnamefont {A.~A.}\ \bibnamefont
  {Kananenka}}, \bibinfo {author} {\bibfnamefont {C.-Y.}\ \bibnamefont
  {Hsieh}}, \bibinfo {author} {\bibfnamefont {J.}~\bibnamefont {Cao}}, \ and\
  \bibinfo {author} {\bibfnamefont {E.}~\bibnamefont {Geva}},\ }\href@noop {}
  {\bibfield  {journal} {\bibinfo  {journal} {J. Phys. Chem. Lett.}\ }\textbf
  {\bibinfo {volume} {7}},\ \bibinfo {pages} {4809} (\bibinfo {year}
  {2016})}\BibitemShut {NoStop}%
\bibitem [{\citenamefont {Gelzinis}\ \emph {et~al.}(2017)\citenamefont
  {Gelzinis}, \citenamefont {Rybakovas},\ and\ \citenamefont
  {Valkunas}}]{gelzinis_applicability_2017}%
  \BibitemOpen
  \bibfield  {author} {\bibinfo {author} {\bibfnamefont {A.}~\bibnamefont
  {Gelzinis}}, \bibinfo {author} {\bibfnamefont {E.}~\bibnamefont {Rybakovas}},
  \ and\ \bibinfo {author} {\bibfnamefont {L.}~\bibnamefont {Valkunas}},\
  }\href@noop {} {\bibfield  {journal} {\bibinfo  {journal} {J. Chem. Phys.}\
  }\textbf {\bibinfo {volume} {147}},\ \bibinfo {pages} {234108} (\bibinfo
  {year} {2017})}\BibitemShut {NoStop}%
\bibitem [{\citenamefont {Pollock}\ and\ \citenamefont
  {Modi}(2018)}]{pollock_tomographically_2018}%
  \BibitemOpen
  \bibfield  {author} {\bibinfo {author} {\bibfnamefont {F.~A.}\ \bibnamefont
  {Pollock}}\ and\ \bibinfo {author} {\bibfnamefont {K.}~\bibnamefont {Modi}},\
  }\href@noop {} {\bibfield  {journal} {\bibinfo  {journal} {Quantum}\ }\textbf
  {\bibinfo {volume} {2}} (\bibinfo {year} {2018})}\BibitemShut {NoStop}%
\bibitem [{\citenamefont {Prokof'ev}\ and\ \citenamefont
  {Stamp}(2000)}]{prokof_stamp_rpp00}%
  \BibitemOpen
  \bibfield  {author} {\bibinfo {author} {\bibfnamefont {N.}~\bibnamefont
  {Prokof'ev}}\ and\ \bibinfo {author} {\bibfnamefont {P.}~\bibnamefont
  {Stamp}},\ }\href@noop {} {\bibfield  {journal} {\bibinfo  {journal} {Rep.
  Prog. Phys.}\ }\textbf {\bibinfo {volume} {63}},\ \bibinfo {pages} {669}
  (\bibinfo {year} {2000})}\BibitemShut {NoStop}%
\bibitem [{\citenamefont {Ma}\ \emph {et~al.}(2015)\citenamefont {Ma},
  \citenamefont {Wolfowicz}, \citenamefont {Li}, \citenamefont {Morton},\ and\
  \citenamefont {Liu}}]{ma_liu_prb15}%
  \BibitemOpen
  \bibfield  {author} {\bibinfo {author} {\bibfnamefont {W.-L.}\ \bibnamefont
  {Ma}}, \bibinfo {author} {\bibfnamefont {G.}~\bibnamefont {Wolfowicz}},
  \bibinfo {author} {\bibfnamefont {S.-S.}\ \bibnamefont {Li}}, \bibinfo
  {author} {\bibfnamefont {J.~J.~L.}\ \bibnamefont {Morton}}, \ and\ \bibinfo
  {author} {\bibfnamefont {R.-B.}\ \bibnamefont {Liu}},\ }\href {\doibase
  10.1103/PhysRevB.92.161403} {\bibfield  {journal} {\bibinfo  {journal} {Phys.
  Rev. B}\ }\textbf {\bibinfo {volume} {92}},\ \bibinfo {pages} {161403}
  (\bibinfo {year} {2015})}\BibitemShut {NoStop}%
\bibitem [{\citenamefont {Hsieh}\ \emph {et~al.}(2012)\citenamefont {Hsieh},
  \citenamefont {Shim}, \citenamefont {Korkusinski},\ and\ \citenamefont
  {Hawrylak}}]{Hsieh_Shim_RPP201}%
  \BibitemOpen
  \bibfield  {author} {\bibinfo {author} {\bibfnamefont {C.-Y.}\ \bibnamefont
  {Hsieh}}, \bibinfo {author} {\bibfnamefont {Y.-P.}\ \bibnamefont {Shim}},
  \bibinfo {author} {\bibfnamefont {M.}~\bibnamefont {Korkusinski}}, \ and\
  \bibinfo {author} {\bibfnamefont {P.}~\bibnamefont {Hawrylak}},\ }\href@noop
  {} {\bibfield  {journal} {\bibinfo  {journal} {Rep. Prog. Phys.}\ }\textbf
  {\bibinfo {volume} {75}},\ \bibinfo {pages} {114501} (\bibinfo {year}
  {2012})}\BibitemShut {NoStop}%
\bibitem [{\citenamefont {Hsieh}\ and\ \citenamefont
  {Cao}(2018)}]{Hsieh_Cao_JCP2018}%
  \BibitemOpen
  \bibfield  {author} {\bibinfo {author} {\bibfnamefont {C.-Y.}\ \bibnamefont
  {Hsieh}}\ and\ \bibinfo {author} {\bibfnamefont {J.}~\bibnamefont {Cao}},\
  }\href@noop {} {\bibfield  {journal} {\bibinfo  {journal} {J. Chem. Phys.}\
  }\textbf {\bibinfo {volume} {148}},\ \bibinfo {pages} {014104} (\bibinfo
  {year} {2018})}\BibitemShut {NoStop}%
\bibitem [{\citenamefont {Merkulov}\ \emph {et~al.}(2002)\citenamefont
  {Merkulov}, \citenamefont {Efros},\ and\ \citenamefont
  {Rosen}}]{merkulov_efros_prb02}%
  \BibitemOpen
  \bibfield  {author} {\bibinfo {author} {\bibfnamefont {I.~A.}\ \bibnamefont
  {Merkulov}}, \bibinfo {author} {\bibfnamefont {A.~L.}\ \bibnamefont {Efros}},
  \ and\ \bibinfo {author} {\bibfnamefont {M.}~\bibnamefont {Rosen}},\
  }\href@noop {} {\bibfield  {journal} {\bibinfo  {journal} {Phys. Rev. B}\
  }\textbf {\bibinfo {volume} {65}},\ \bibinfo {pages} {205309} (\bibinfo
  {year} {2002})}\BibitemShut {NoStop}%
\bibitem [{\citenamefont {Erlingsson}\ and\ \citenamefont
  {Nazarov}(2002)}]{erlingsson_hyperfine_2002}%
  \BibitemOpen
  \bibfield  {author} {\bibinfo {author} {\bibfnamefont {S.~I.}\ \bibnamefont
  {Erlingsson}}\ and\ \bibinfo {author} {\bibfnamefont {Y.~V.}\ \bibnamefont
  {Nazarov}},\ }\href@noop {} {\bibfield  {journal} {\bibinfo  {journal} {Phys.
  Rev. B}\ }\textbf {\bibinfo {volume} {66}} (\bibinfo {year}
  {2002})}\BibitemShut {NoStop}%
\bibitem [{\citenamefont {Witzel}\ \emph {et~al.}(2014)\citenamefont {Witzel},
  \citenamefont {Young},\ and\ \citenamefont {Das~Sarma}}]{witzel_sarma_prb14}%
  \BibitemOpen
  \bibfield  {author} {\bibinfo {author} {\bibfnamefont {W.~M.}\ \bibnamefont
  {Witzel}}, \bibinfo {author} {\bibfnamefont {K.}~\bibnamefont {Young}}, \
  and\ \bibinfo {author} {\bibfnamefont {S.}~\bibnamefont {Das~Sarma}},\ }\href
  {\doibase 10.1103/PhysRevB.90.115431} {\bibfield  {journal} {\bibinfo
  {journal} {Phys. Rev. B}\ }\textbf {\bibinfo {volume} {90}},\ \bibinfo
  {pages} {115431} (\bibinfo {year} {2014})}\BibitemShut {NoStop}%
\bibitem [{\citenamefont {Makri}(1999)}]{makri_jpcb1999}%
  \BibitemOpen
  \bibfield  {author} {\bibinfo {author} {\bibfnamefont {N.}~\bibnamefont
  {Makri}},\ }\href@noop {} {\bibfield  {journal} {\bibinfo  {journal} {J.
  Phys. Chem. B}\ }\textbf {\bibinfo {volume} {103}},\ \bibinfo {pages} {2823}
  (\bibinfo {year} {1999})}\BibitemShut {NoStop}%
\bibitem [{\citenamefont {Caldeira}\ and\ \citenamefont
  {Leggett}(1983)}]{caldeira_leggett_annals83}%
  \BibitemOpen
  \bibfield  {author} {\bibinfo {author} {\bibfnamefont {A.}~\bibnamefont
  {Caldeira}}\ and\ \bibinfo {author} {\bibfnamefont {A.~J.}\ \bibnamefont
  {Leggett}},\ }\href@noop {} {\bibfield  {journal} {\bibinfo  {journal} {Ann.
  Phys.}\ }\textbf {\bibinfo {volume} {149}},\ \bibinfo {pages} {374} (\bibinfo
  {year} {1983})}\BibitemShut {NoStop}%
\bibitem [{\citenamefont {Viola}\ and\ \citenamefont
  {Lloyd}(1998)}]{PhysRevA.58.2733}%
  \BibitemOpen
  \bibfield  {author} {\bibinfo {author} {\bibfnamefont {L.}~\bibnamefont
  {Viola}}\ and\ \bibinfo {author} {\bibfnamefont {S.}~\bibnamefont {Lloyd}},\
  }\href@noop {} {\bibfield  {journal} {\bibinfo  {journal} {Phys. Rev. A}\
  }\textbf {\bibinfo {volume} {58}},\ \bibinfo {pages} {2733} (\bibinfo {year}
  {1998})}\BibitemShut {NoStop}%
\bibitem [{\citenamefont {Khodjasteh}\ and\ \citenamefont
  {Lidar}(2005)}]{khodjasteh_lidar_prl05}%
  \BibitemOpen
  \bibfield  {author} {\bibinfo {author} {\bibfnamefont {K.}~\bibnamefont
  {Khodjasteh}}\ and\ \bibinfo {author} {\bibfnamefont {D.}~\bibnamefont
  {Lidar}},\ }\href@noop {} {\bibfield  {journal} {\bibinfo  {journal} {Phys.
  Rev. Lett.}\ }\textbf {\bibinfo {volume} {95}},\ \bibinfo {pages} {180501}
  (\bibinfo {year} {2005})}\BibitemShut {NoStop}%
\bibitem [{\citenamefont {Khodjasteh}\ \emph {et~al.}(2010)\citenamefont
  {Khodjasteh}, \citenamefont {Lidar},\ and\ \citenamefont
  {Viola}}]{khodjasteh_lidar_prl10}%
  \BibitemOpen
  \bibfield  {author} {\bibinfo {author} {\bibfnamefont {K.}~\bibnamefont
  {Khodjasteh}}, \bibinfo {author} {\bibfnamefont {D.~A.}\ \bibnamefont
  {Lidar}}, \ and\ \bibinfo {author} {\bibfnamefont {L.}~\bibnamefont
  {Viola}},\ }\href@noop {} {\bibfield  {journal} {\bibinfo  {journal} {Phys.
  Rev. Lett.}\ }\textbf {\bibinfo {volume} {104}},\ \bibinfo {pages} {090501}
  (\bibinfo {year} {2010})}\BibitemShut {NoStop}%
\bibitem [{\citenamefont {Yang}\ \emph {et~al.}(2011)\citenamefont {Yang},
  \citenamefont {Wang},\ and\ \citenamefont {Liu}}]{yang_liu_frontphys11}%
  \BibitemOpen
  \bibfield  {author} {\bibinfo {author} {\bibfnamefont {W.}~\bibnamefont
  {Yang}}, \bibinfo {author} {\bibfnamefont {Z.-Y.}\ \bibnamefont {Wang}}, \
  and\ \bibinfo {author} {\bibfnamefont {R.-B.}\ \bibnamefont {Liu}},\
  }\href@noop {} {\bibfield  {journal} {\bibinfo  {journal} {Front. Phys.}\
  }\textbf {\bibinfo {volume} {6}},\ \bibinfo {pages} {2} (\bibinfo {year}
  {2011})}\BibitemShut {NoStop}%
\bibitem [{\citenamefont {Szankowski}\ \emph {et~al.}(2017)\citenamefont
  {Szankowski}, \citenamefont {Ramon}, \citenamefont {Krzywda}, \citenamefont
  {Kwiatkowski},\ and\ \citenamefont
  {Cywinski}}]{szankowski_environmental_2017}%
  \BibitemOpen
  \bibfield  {author} {\bibinfo {author} {\bibfnamefont {P.}~\bibnamefont
  {Szankowski}}, \bibinfo {author} {\bibfnamefont {G.}~\bibnamefont {Ramon}},
  \bibinfo {author} {\bibfnamefont {J.}~\bibnamefont {Krzywda}}, \bibinfo
  {author} {\bibfnamefont {D.}~\bibnamefont {Kwiatkowski}}, \ and\ \bibinfo
  {author} {\bibfnamefont {L.}~\bibnamefont {Cywinski}},\ }\href@noop {}
  {\bibfield  {journal} {\bibinfo  {journal} {J. Phys.: Condens. Matter}\
  }\textbf {\bibinfo {volume} {29}},\ \bibinfo {pages} {333001} (\bibinfo
  {year} {2017})}\BibitemShut {NoStop}%
\bibitem [{\citenamefont {√Ålvarez}\ and\ \citenamefont
  {Suter}(2011)}]{alvarez_measuring_2011}%
  \BibitemOpen
  \bibfield  {author} {\bibinfo {author} {\bibfnamefont {G.~A.}\ \bibnamefont
  {√Ålvarez}}\ and\ \bibinfo {author} {\bibfnamefont {D.}~\bibnamefont
  {Suter}},\ }\href@noop {} {\bibfield  {journal} {\bibinfo  {journal} {Phys.
  Rev. Lett.}\ }\textbf {\bibinfo {volume} {107}},\ \bibinfo {pages} {230501}
  (\bibinfo {year} {2011})}\BibitemShut {NoStop}%
\bibitem [{\citenamefont {Zwick}\ \emph {et~al.}(2016)\citenamefont {Zwick},
  \citenamefont {Alvarez},\ and\ \citenamefont
  {Kurizki}}]{zwick_maximizing_2016}%
  \BibitemOpen
  \bibfield  {author} {\bibinfo {author} {\bibfnamefont {A.}~\bibnamefont
  {Zwick}}, \bibinfo {author} {\bibfnamefont {G.~A.}\ \bibnamefont {Alvarez}},
  \ and\ \bibinfo {author} {\bibfnamefont {G.}~\bibnamefont {Kurizki}},\
  }\href@noop {} {\bibfield  {journal} {\bibinfo  {journal} {Phys. Rev. Appl.}\
  }\textbf {\bibinfo {volume} {5}},\ \bibinfo {pages} {014007} (\bibinfo {year}
  {2016})}\BibitemShut {NoStop}%
\bibitem [{\citenamefont {Norris}\ \emph {et~al.}(2016)\citenamefont {Norris},
  \citenamefont {Paz-Silva},\ and\ \citenamefont {Viola}}]{norris_qubit_2016}%
  \BibitemOpen
  \bibfield  {author} {\bibinfo {author} {\bibfnamefont {L.~M.}\ \bibnamefont
  {Norris}}, \bibinfo {author} {\bibfnamefont {G.~A.}\ \bibnamefont
  {Paz-Silva}}, \ and\ \bibinfo {author} {\bibfnamefont {L.}~\bibnamefont
  {Viola}},\ }\href@noop {} {\bibfield  {journal} {\bibinfo  {journal} {Phys.
  Rev. Lett.}\ }\textbf {\bibinfo {volume} {116}},\ \bibinfo {pages} {150503}
  (\bibinfo {year} {2016})}\BibitemShut {NoStop}%
\bibitem [{\citenamefont {Paz-Silva}\ \emph {et~al.}(2017)\citenamefont
  {Paz-Silva}, \citenamefont {Norris},\ and\ \citenamefont
  {Viola}}]{paz-silva_multiqubit_2017}%
  \BibitemOpen
  \bibfield  {author} {\bibinfo {author} {\bibfnamefont {G.~A.}\ \bibnamefont
  {Paz-Silva}}, \bibinfo {author} {\bibfnamefont {L.~M.}\ \bibnamefont
  {Norris}}, \ and\ \bibinfo {author} {\bibfnamefont {L.}~\bibnamefont
  {Viola}},\ }\href@noop {} {\bibfield  {journal} {\bibinfo  {journal} {Phys.
  Rev. A}\ }\textbf {\bibinfo {volume} {95}},\ \bibinfo {pages} {022121}
  (\bibinfo {year} {2017})}\BibitemShut {NoStop}%
\bibitem [{\citenamefont {Krzywda}\ \emph {et~al.}(2018)\citenamefont
  {Krzywda}, \citenamefont {Sza≈Ñkowski},\ and\ \citenamefont
  {Cywi≈Ñski}}]{krzywda_dynamical-decoupling-based_2018}%
  \BibitemOpen
  \bibfield  {author} {\bibinfo {author} {\bibfnamefont {J.}~\bibnamefont
  {Krzywda}}, \bibinfo {author} {\bibfnamefont {P.}~\bibnamefont
  {Sza≈Ñkowski}}, \ and\ \bibinfo {author} {\bibfnamefont {≈.}~\bibnamefont
  {Cywi≈Ñski}},\ }\href@noop {} {\bibfield  {journal} {\bibinfo  {journal}
  {arXiv:1809.02972}\ } (\bibinfo {year} {2018})}\BibitemShut {NoStop}%
\bibitem [{\citenamefont {Cywinski}(2014)}]{cywinski_dynamical_2014}%
  \BibitemOpen
  \bibfield  {author} {\bibinfo {author} {\bibfnamefont {L.}~\bibnamefont
  {Cywinski}},\ }\href@noop {} {\bibfield  {journal} {\bibinfo  {journal}
  {Phys. Rev. A}\ }\textbf {\bibinfo {volume} {90}},\ \bibinfo {pages} {042307}
  (\bibinfo {year} {2014})}\BibitemShut {NoStop}%
\bibitem [{\citenamefont {Yuge}\ \emph {et~al.}(2011)\citenamefont {Yuge},
  \citenamefont {Sasaki},\ and\ \citenamefont
  {Hirayama}}]{yuge_measurement_2011}%
  \BibitemOpen
  \bibfield  {author} {\bibinfo {author} {\bibfnamefont {T.}~\bibnamefont
  {Yuge}}, \bibinfo {author} {\bibfnamefont {S.}~\bibnamefont {Sasaki}}, \ and\
  \bibinfo {author} {\bibfnamefont {Y.}~\bibnamefont {Hirayama}},\ }\href@noop
  {} {\bibfield  {journal} {\bibinfo  {journal} {Phys. Rev. Lett.}\ }\textbf
  {\bibinfo {volume} {107}},\ \bibinfo {pages} {170504} (\bibinfo {year}
  {2011})}\BibitemShut {NoStop}%
\bibitem [{\citenamefont {Ma}\ and\ \citenamefont
  {Liu}(2016)}]{ma_proposal_2016}%
  \BibitemOpen
  \bibfield  {author} {\bibinfo {author} {\bibfnamefont {W.-L.}\ \bibnamefont
  {Ma}}\ and\ \bibinfo {author} {\bibfnamefont {R.-B.}\ \bibnamefont {Liu}},\
  }\href@noop {} {\bibfield  {journal} {\bibinfo  {journal} {Phys. Rev. Appl.}\
  }\textbf {\bibinfo {volume} {6}},\ \bibinfo {pages} {054012} (\bibinfo {year}
  {2016})}\BibitemShut {NoStop}%
\bibitem [{\citenamefont {Paik}\ \emph {et~al.}(2011)\citenamefont {Paik},
  \citenamefont {Schuster}, \citenamefont {Bishop}, \citenamefont {Kirchmair},
  \citenamefont {Catelani}, \citenamefont {Sears}, \citenamefont {Johnson},
  \citenamefont {Reagor}, \citenamefont {Frunzio}, \citenamefont {Glazman},
  \citenamefont {Girvin}, \citenamefont {Devoret},\ and\ \citenamefont
  {Schoelkopf}}]{hanhee_schuster_prl11}%
  \BibitemOpen
  \bibfield  {author} {\bibinfo {author} {\bibfnamefont {H.}~\bibnamefont
  {Paik}}, \bibinfo {author} {\bibfnamefont {D.~I.}\ \bibnamefont {Schuster}},
  \bibinfo {author} {\bibfnamefont {L.~S.}\ \bibnamefont {Bishop}}, \bibinfo
  {author} {\bibfnamefont {G.}~\bibnamefont {Kirchmair}}, \bibinfo {author}
  {\bibfnamefont {G.}~\bibnamefont {Catelani}}, \bibinfo {author}
  {\bibfnamefont {A.~P.}\ \bibnamefont {Sears}}, \bibinfo {author}
  {\bibfnamefont {B.~R.}\ \bibnamefont {Johnson}}, \bibinfo {author}
  {\bibfnamefont {M.~J.}\ \bibnamefont {Reagor}}, \bibinfo {author}
  {\bibfnamefont {L.}~\bibnamefont {Frunzio}}, \bibinfo {author} {\bibfnamefont
  {L.~I.}\ \bibnamefont {Glazman}}, \bibinfo {author} {\bibfnamefont {S.~M.}\
  \bibnamefont {Girvin}}, \bibinfo {author} {\bibfnamefont {M.~H.}\
  \bibnamefont {Devoret}}, \ and\ \bibinfo {author} {\bibfnamefont {R.~J.}\
  \bibnamefont {Schoelkopf}},\ }\href@noop {} {\bibfield  {journal} {\bibinfo
  {journal} {Phys. Rev. Lett.}\ }\textbf {\bibinfo {volume} {107}},\ \bibinfo
  {pages} {240501} (\bibinfo {year} {2011})}\BibitemShut {NoStop}%
\bibitem [{\citenamefont {Rigetti}\ \emph {et~al.}(2012)\citenamefont
  {Rigetti}, \citenamefont {Gambetta}, \citenamefont {Poletto}, \citenamefont
  {Plourde}, \citenamefont {Chow}, \citenamefont {C{\'o}rcoles}, \citenamefont
  {Smolin}, \citenamefont {Merkel}, \citenamefont {Rozen}, \citenamefont
  {Keefe} \emph {et~al.}}]{rigetti_gambetta_prb12}%
  \BibitemOpen
  \bibfield  {author} {\bibinfo {author} {\bibfnamefont {C.}~\bibnamefont
  {Rigetti}}, \bibinfo {author} {\bibfnamefont {J.~M.}\ \bibnamefont
  {Gambetta}}, \bibinfo {author} {\bibfnamefont {S.}~\bibnamefont {Poletto}},
  \bibinfo {author} {\bibfnamefont {B.}~\bibnamefont {Plourde}}, \bibinfo
  {author} {\bibfnamefont {J.~M.}\ \bibnamefont {Chow}}, \bibinfo {author}
  {\bibfnamefont {A.}~\bibnamefont {C{\'o}rcoles}}, \bibinfo {author}
  {\bibfnamefont {J.~A.}\ \bibnamefont {Smolin}}, \bibinfo {author}
  {\bibfnamefont {S.~T.}\ \bibnamefont {Merkel}}, \bibinfo {author}
  {\bibfnamefont {J.}~\bibnamefont {Rozen}}, \bibinfo {author} {\bibfnamefont
  {G.~A.}\ \bibnamefont {Keefe}},  \emph {et~al.},\ }\href@noop {} {\bibfield
  {journal} {\bibinfo  {journal} {Phys. Rev. B}\ }\textbf {\bibinfo {volume}
  {86}},\ \bibinfo {pages} {100506} (\bibinfo {year} {2012})}\BibitemShut
  {NoStop}%
\bibitem [{\citenamefont {Chen}\ \emph {et~al.}(2014)\citenamefont {Chen},
  \citenamefont {Neill}, \citenamefont {Roushan}, \citenamefont {Leung},
  \citenamefont {Fang}, \citenamefont {Barends}, \citenamefont {Kelly},
  \citenamefont {Campbell}, \citenamefont {Chen}, \citenamefont {Chiaro},
  \citenamefont {Dunsworth}, \citenamefont {Jeffrey}, \citenamefont {Megrant},
  \citenamefont {Mutus}, \citenamefont {O'Malley}, \citenamefont {Quintana},
  \citenamefont {Sank}, \citenamefont {Vainsencher}, \citenamefont {Wenner},
  \citenamefont {White}, \citenamefont {Geller}, \citenamefont {Cleland},\ and\
  \citenamefont {Martinis}}]{chen2014qubit}%
  \BibitemOpen
  \bibfield  {author} {\bibinfo {author} {\bibfnamefont {Y.}~\bibnamefont
  {Chen}}, \bibinfo {author} {\bibfnamefont {C.}~\bibnamefont {Neill}},
  \bibinfo {author} {\bibfnamefont {P.}~\bibnamefont {Roushan}}, \bibinfo
  {author} {\bibfnamefont {N.}~\bibnamefont {Leung}}, \bibinfo {author}
  {\bibfnamefont {M.}~\bibnamefont {Fang}}, \bibinfo {author} {\bibfnamefont
  {R.}~\bibnamefont {Barends}}, \bibinfo {author} {\bibfnamefont
  {J.}~\bibnamefont {Kelly}}, \bibinfo {author} {\bibfnamefont
  {B.}~\bibnamefont {Campbell}}, \bibinfo {author} {\bibfnamefont
  {Z.}~\bibnamefont {Chen}}, \bibinfo {author} {\bibfnamefont {B.}~\bibnamefont
  {Chiaro}}, \bibinfo {author} {\bibfnamefont {A.}~\bibnamefont {Dunsworth}},
  \bibinfo {author} {\bibfnamefont {E.}~\bibnamefont {Jeffrey}}, \bibinfo
  {author} {\bibfnamefont {A.}~\bibnamefont {Megrant}}, \bibinfo {author}
  {\bibfnamefont {J.~Y.}\ \bibnamefont {Mutus}}, \bibinfo {author}
  {\bibfnamefont {P.~J.~J.}\ \bibnamefont {O'Malley}}, \bibinfo {author}
  {\bibfnamefont {C.~M.}\ \bibnamefont {Quintana}}, \bibinfo {author}
  {\bibfnamefont {D.}~\bibnamefont {Sank}}, \bibinfo {author} {\bibfnamefont
  {A.}~\bibnamefont {Vainsencher}}, \bibinfo {author} {\bibfnamefont
  {J.}~\bibnamefont {Wenner}}, \bibinfo {author} {\bibfnamefont {T.~C.}\
  \bibnamefont {White}}, \bibinfo {author} {\bibfnamefont {M.~R.}\ \bibnamefont
  {Geller}}, \bibinfo {author} {\bibfnamefont {A.~N.}\ \bibnamefont {Cleland}},
  \ and\ \bibinfo {author} {\bibfnamefont {J.~M.}\ \bibnamefont {Martinis}},\
  }\href@noop {} {\bibfield  {journal} {\bibinfo  {journal} {Phys. Rev. Lett.}\
  }\textbf {\bibinfo {volume} {113}},\ \bibinfo {pages} {220502} (\bibinfo
  {year} {2014})}\BibitemShut {NoStop}%
\bibitem [{\citenamefont {Shiokawa}\ and\ \citenamefont
  {Lidar}(2004)}]{shiokawa_lidar_pra04}%
  \BibitemOpen
  \bibfield  {author} {\bibinfo {author} {\bibfnamefont {K.}~\bibnamefont
  {Shiokawa}}\ and\ \bibinfo {author} {\bibfnamefont {D.}~\bibnamefont
  {Lidar}},\ }\href@noop {} {\bibfield  {journal} {\bibinfo  {journal} {Phys.
  Rev. A}\ }\textbf {\bibinfo {volume} {69}},\ \bibinfo {pages} {030302}
  (\bibinfo {year} {2004})}\BibitemShut {NoStop}%
\bibitem [{\citenamefont {Lidar}\ and\ \citenamefont
  {Brun}(2013)}]{lidar_quantum_2013}%
  \BibitemOpen
  \bibinfo {editor} {\bibfnamefont {D.~A.}\ \bibnamefont {Lidar}}\ and\
  \bibinfo {editor} {\bibfnamefont {T.~A.}\ \bibnamefont {Brun}},\ eds.,\
  \href@noop {} {{\emph {\bibinfo {title} {Quantum error correction}}}}\ (\bibinfo  {publisher} {Cambridge University Press},\
  \bibinfo {address} {Cambridge, United Kingdom ; New York},\ \bibinfo {year}
  {2013})\BibitemShut {NoStop}%
\bibitem [{\citenamefont {Gross}\ \emph {et~al.}(2010)\citenamefont {Gross},
  \citenamefont {Liu}, \citenamefont {Flammia}, \citenamefont {Becker},\ and\
  \citenamefont {Jens}}]{gross_liu_prl10}%
  \BibitemOpen
  \bibfield  {author} {\bibinfo {author} {\bibfnamefont {D.}~\bibnamefont
  {Gross}}, \bibinfo {author} {\bibfnamefont {Y.-K.}\ \bibnamefont {Liu}},
  \bibinfo {author} {\bibfnamefont {S.~T.}\ \bibnamefont {Flammia}}, \bibinfo
  {author} {\bibfnamefont {S.}~\bibnamefont {Becker}}, \ and\ \bibinfo {author}
  {\bibfnamefont {E.}~\bibnamefont {Jens}},\ }\href@noop {} {\bibfield
  {journal} {\bibinfo  {journal} {Phys. Rev. Lett.}\ }\textbf {\bibinfo
  {volume} {105}},\ \bibinfo {pages} {150401} (\bibinfo {year}
  {2010})}\BibitemShut {NoStop}%
\bibitem [{\citenamefont {Riofr√≠o}\ \emph {et~al.}(2017)\citenamefont
  {Riofr√≠o}, \citenamefont {Gross}, \citenamefont {Flammia}, \citenamefont
  {Monz}, \citenamefont {Nigg}, \citenamefont {Blatt},\ and\ \citenamefont
  {Eisert}}]{riofrio_gross_natcomm17}%
  \BibitemOpen
  \bibfield  {author} {\bibinfo {author} {\bibfnamefont {C.~A.}\ \bibnamefont
  {Riofr√≠o}}, \bibinfo {author} {\bibfnamefont {D.}~\bibnamefont {Gross}},
  \bibinfo {author} {\bibfnamefont {S.~T.}\ \bibnamefont {Flammia}}, \bibinfo
  {author} {\bibfnamefont {T.}~\bibnamefont {Monz}}, \bibinfo {author}
  {\bibfnamefont {D.}~\bibnamefont {Nigg}}, \bibinfo {author} {\bibfnamefont
  {R.}~\bibnamefont {Blatt}}, \ and\ \bibinfo {author} {\bibfnamefont
  {J.}~\bibnamefont {Eisert}},\ }\href@noop {} {\bibfield  {journal} {\bibinfo
  {journal} {Nat. Comm.}\ }\textbf {\bibinfo {volume} {8}},\ \bibinfo {pages}
  {15305} (\bibinfo {year} {2017})}\BibitemShut {NoStop}%
\bibitem [{\citenamefont {Torlai}\ \emph {et~al.}(2018)\citenamefont {Torlai},
  \citenamefont {Mazzola}, \citenamefont {Carrasquilla}, \citenamefont
  {Troyer}, \citenamefont {Melko},\ and\ \citenamefont
  {Carleo}}]{torlai_mazzola_natphys18}%
  \BibitemOpen
  \bibfield  {author} {\bibinfo {author} {\bibfnamefont {G.}~\bibnamefont
  {Torlai}}, \bibinfo {author} {\bibfnamefont {G.}~\bibnamefont {Mazzola}},
  \bibinfo {author} {\bibfnamefont {J.}~\bibnamefont {Carrasquilla}}, \bibinfo
  {author} {\bibfnamefont {M.}~\bibnamefont {Troyer}}, \bibinfo {author}
  {\bibfnamefont {R.}~\bibnamefont {Melko}}, \ and\ \bibinfo {author}
  {\bibfnamefont {G.}~\bibnamefont {Carleo}},\ }\href@noop {} {\bibfield
  {journal} {\bibinfo  {journal} {Nat. Phys.}\ }\textbf {\bibinfo {volume}
  {14}},\ \bibinfo {pages} {447‚Äì450} (\bibinfo {year} {2018})}\BibitemShut
  {NoStop}%
\bibitem [{\citenamefont {Rocchetto}\ \emph {et~al.}(2018)\citenamefont
  {Rocchetto}, \citenamefont {Grant}, \citenamefont {Strelchuk}, \citenamefont
  {Carleo},\ and\ \citenamefont {Severini}}]{rochcetto_grant_npjqip18}%
  \BibitemOpen
  \bibfield  {author} {\bibinfo {author} {\bibfnamefont {A.}~\bibnamefont
  {Rocchetto}}, \bibinfo {author} {\bibfnamefont {E.}~\bibnamefont {Grant}},
  \bibinfo {author} {\bibfnamefont {S.}~\bibnamefont {Strelchuk}}, \bibinfo
  {author} {\bibfnamefont {G.}~\bibnamefont {Carleo}}, \ and\ \bibinfo {author}
  {\bibfnamefont {S.}~\bibnamefont {Severini}},\ }\href@noop {} {\bibfield
  {journal} {\bibinfo  {journal} {npj Quantum Inf.}\ }\textbf {\bibinfo
  {volume} {4}},\ \bibinfo {pages} {28} (\bibinfo {year} {2018})}\BibitemShut
  {NoStop}%
\bibitem [{\citenamefont {Torlai}\ and\ \citenamefont
  {Melko}(2018)}]{giacomo_melko_arxiv18}%
  \BibitemOpen
  \bibfield  {author} {\bibinfo {author} {\bibfnamefont {G.}~\bibnamefont
  {Torlai}}\ and\ \bibinfo {author} {\bibfnamefont {R.~G.}\ \bibnamefont
  {Melko}},\ }\href@noop {} {\bibfield  {journal} {\bibinfo  {journal}
  {arxiv:1801.09684}\ } (\bibinfo {year} {2018})}\BibitemShut {NoStop}%
\bibitem [{\citenamefont {Cramer}\ \emph {et~al.}(2010)\citenamefont {Cramer},
  \citenamefont {Plenio}, \citenamefont {Flammia}, \citenamefont {Somma},
  \citenamefont {Gross}, \citenamefont {Bartlett}, \citenamefont
  {Landon-Cardinal}, \citenamefont {Poulin},\ and\ \citenamefont
  {Liu}}]{cramer_plenio_nat2010}%
  \BibitemOpen
  \bibfield  {author} {\bibinfo {author} {\bibfnamefont {M.}~\bibnamefont
  {Cramer}}, \bibinfo {author} {\bibfnamefont {M.~B.}\ \bibnamefont {Plenio}},
  \bibinfo {author} {\bibfnamefont {S.~T.}\ \bibnamefont {Flammia}}, \bibinfo
  {author} {\bibfnamefont {R.}~\bibnamefont {Somma}}, \bibinfo {author}
  {\bibfnamefont {D.}~\bibnamefont {Gross}}, \bibinfo {author} {\bibfnamefont
  {S.~D.}\ \bibnamefont {Bartlett}}, \bibinfo {author} {\bibfnamefont
  {O.}~\bibnamefont {Landon-Cardinal}}, \bibinfo {author} {\bibfnamefont
  {D.}~\bibnamefont {Poulin}}, \ and\ \bibinfo {author} {\bibfnamefont {Y.~K.}\
  \bibnamefont {Liu}},\ }\href@noop {} {\bibfield  {journal} {\bibinfo
  {journal} {Nat. Comm.}\ }\textbf {\bibinfo {volume} {1}},\ \bibinfo {pages}
  {149} (\bibinfo {year} {2010})}\BibitemShut {NoStop}%
\bibitem [{\citenamefont {Lanyon}\ \emph {et~al.}(2017)\citenamefont {Lanyon},
  \citenamefont {Maier}, \citenamefont {Holz√§pfel}, \citenamefont {Baumgratz},
  \citenamefont {Hempel}, \citenamefont {Jurcevic}, \citenamefont {Dhand},
  \citenamefont {Buyskikh}, \citenamefont {Daley}, \citenamefont {Cramer},
  \citenamefont {Plenio}, \citenamefont {Blatt},\ and\ \citenamefont
  {Roos}}]{lanyon_maier_natphys17}%
  \BibitemOpen
  \bibfield  {author} {\bibinfo {author} {\bibfnamefont {B.~P.}\ \bibnamefont
  {Lanyon}}, \bibinfo {author} {\bibfnamefont {C.}~\bibnamefont {Maier}},
  \bibinfo {author} {\bibfnamefont {M.}~\bibnamefont {Holz√§pfel}}, \bibinfo
  {author} {\bibfnamefont {T.}~\bibnamefont {Baumgratz}}, \bibinfo {author}
  {\bibfnamefont {C.}~\bibnamefont {Hempel}}, \bibinfo {author} {\bibfnamefont
  {P.}~\bibnamefont {Jurcevic}}, \bibinfo {author} {\bibfnamefont
  {I.}~\bibnamefont {Dhand}}, \bibinfo {author} {\bibfnamefont {A.~S.}\
  \bibnamefont {Buyskikh}}, \bibinfo {author} {\bibfnamefont {A.~J.}\
  \bibnamefont {Daley}}, \bibinfo {author} {\bibfnamefont {M.}~\bibnamefont
  {Cramer}}, \bibinfo {author} {\bibfnamefont {M.~B.}\ \bibnamefont {Plenio}},
  \bibinfo {author} {\bibfnamefont {R.}~\bibnamefont {Blatt}}, \ and\ \bibinfo
  {author} {\bibfnamefont {C.~F.}\ \bibnamefont {Roos}},\ }\href@noop {}
  {\bibfield  {journal} {\bibinfo  {journal} {Nat. Phys.}\ }\textbf {\bibinfo
  {volume} {13}},\ \bibinfo {pages} {1158} (\bibinfo {year}
  {2017})}\BibitemShut {NoStop}%
\bibitem [{\citenamefont {Palmieri}\ \emph {et~al.}()\citenamefont {Palmieri},
  \citenamefont {Kovlakov}, \citenamefont {Bianchi}, \citenamefont {Yudin},
  \citenamefont {Straupe}, \citenamefont {Biamonte},\ and\ \citenamefont
  {Kulik}}]{palmieri_kovlakov_arxiv19}%
  \BibitemOpen
  \bibfield  {author} {\bibinfo {author} {\bibfnamefont {A.~M.}\ \bibnamefont
  {Palmieri}}, \bibinfo {author} {\bibfnamefont {E.}~\bibnamefont {Kovlakov}},
  \bibinfo {author} {\bibfnamefont {F.}~\bibnamefont {Bianchi}}, \bibinfo
  {author} {\bibfnamefont {D.}~\bibnamefont {Yudin}}, \bibinfo {author}
  {\bibfnamefont {S.}~\bibnamefont {Straupe}}, \bibinfo {author} {\bibfnamefont
  {J.~D.}\ \bibnamefont {Biamonte}}, \ and\ \bibinfo {author} {\bibfnamefont
  {S.}~\bibnamefont {Kulik}},\ }\href@noop {} {\bibinfo  {journal}
  {arxiv:1904.05902}\ }\BibitemShut {NoStop}%
\bibitem [{\citenamefont {Merkel}\ \emph {et~al.}(2013)\citenamefont {Merkel},
  \citenamefont {Gambetta}, \citenamefont {Smolin}, \citenamefont {Poletto},
  \citenamefont {C\'orcoles}, \citenamefont {Johnson}, \citenamefont {Ryan},\
  and\ \citenamefont {Steffen}}]{seth_gambetta_pra13}%
  \BibitemOpen
\bibfield  {journal} {  }\bibfield  {author} {\bibinfo {author} {\bibfnamefont
  {S.~T.}\ \bibnamefont {Merkel}}, \bibinfo {author} {\bibfnamefont {J.~M.}\
  \bibnamefont {Gambetta}}, \bibinfo {author} {\bibfnamefont {J.~A.}\
  \bibnamefont {Smolin}}, \bibinfo {author} {\bibfnamefont {S.}~\bibnamefont
  {Poletto}}, \bibinfo {author} {\bibfnamefont {A.~D.}\ \bibnamefont
  {C\'orcoles}}, \bibinfo {author} {\bibfnamefont {B.~R.}\ \bibnamefont
  {Johnson}}, \bibinfo {author} {\bibfnamefont {C.~A.}\ \bibnamefont {Ryan}}, \
  and\ \bibinfo {author} {\bibfnamefont {M.}~\bibnamefont {Steffen}},\
  }\href@noop {} {\bibfield  {journal} {\bibinfo  {journal} {Phys. Rev. A}\
  }\textbf {\bibinfo {volume} {87}},\ \bibinfo {pages} {062119} (\bibinfo
  {year} {2013})}\BibitemShut {NoStop}%
\bibitem [{\citenamefont {Blume-Kohout}\ \emph {et~al.}(2013)\citenamefont
  {Blume-Kohout}, \citenamefont {Gamble}, \citenamefont {Nielsen},
  \citenamefont {Rudinger}, \citenamefont {Mizrahi}, \citenamefont {Fortier},\
  and\ \citenamefont {Maunz}}]{kohout_gamble_natcomm13}%
  \BibitemOpen
  \bibfield  {author} {\bibinfo {author} {\bibfnamefont {R.}~\bibnamefont
  {Blume-Kohout}}, \bibinfo {author} {\bibfnamefont {J.~K.}\ \bibnamefont
  {Gamble}}, \bibinfo {author} {\bibfnamefont {E.}~\bibnamefont {Nielsen}},
  \bibinfo {author} {\bibfnamefont {K.}~\bibnamefont {Rudinger}}, \bibinfo
  {author} {\bibfnamefont {J.}~\bibnamefont {Mizrahi}}, \bibinfo {author}
  {\bibfnamefont {K.}~\bibnamefont {Fortier}}, \ and\ \bibinfo {author}
  {\bibfnamefont {P.}~\bibnamefont {Maunz}},\ }\href@noop {} {\bibfield
  {journal} {\bibinfo  {journal} {Nat. Comm.}\ }\textbf {\bibinfo {volume}
  {8}},\ \bibinfo {pages} {14485} (\bibinfo {year} {2013})}\BibitemShut
  {NoStop}%
\bibitem [{\citenamefont {Rivas}\ \emph {et~al.}(2014)\citenamefont {Rivas},
  \citenamefont {Huelga},\ and\ \citenamefont {Plenio}}]{rivas_quantum_2014}%
  \BibitemOpen
  \bibfield  {author} {\bibinfo {author} {\bibfnamefont {A.}~\bibnamefont
  {Rivas}}, \bibinfo {author} {\bibfnamefont {S.~F.}\ \bibnamefont {Huelga}}, \
  and\ \bibinfo {author} {\bibfnamefont {M.~B.}\ \bibnamefont {Plenio}},\
  }\href {\doibase 10.1088/0034-4885/77/9/094001} {\bibfield  {journal}
  {\bibinfo  {journal} {Rep. Prog. Phys.}\ }\textbf {\bibinfo {volume} {77}},\
  \bibinfo {pages} {094001} (\bibinfo {year} {2014})}\BibitemShut {NoStop}%
\bibitem [{\citenamefont {Pollock}\ \emph {et~al.}(2018)\citenamefont
  {Pollock}, \citenamefont {Rodr√≠guez-Rosario}, \citenamefont {Frauenheim},
  \citenamefont {Paternostro},\ and\ \citenamefont
  {Modi}}]{pollock_non-markovian_2018}%
  \BibitemOpen
  \bibfield  {author} {\bibinfo {author} {\bibfnamefont {F.~A.}\ \bibnamefont
  {Pollock}}, \bibinfo {author} {\bibfnamefont {C.}~\bibnamefont
  {Rodr√≠guez-Rosario}}, \bibinfo {author} {\bibfnamefont {T.}~\bibnamefont
  {Frauenheim}}, \bibinfo {author} {\bibfnamefont {M.}~\bibnamefont
  {Paternostro}}, \ and\ \bibinfo {author} {\bibfnamefont {K.}~\bibnamefont
  {Modi}},\ }\href@noop {} {\bibfield  {journal} {\bibinfo  {journal} {Phys.
  Rev. A}\ }\textbf {\bibinfo {volume} {97}},\ \bibinfo {pages} {012127}
  (\bibinfo {year} {2018})}\BibitemShut {NoStop}%
\bibitem [{\citenamefont {Lorenzo}\ \emph {et~al.}(2013)\citenamefont
  {Lorenzo}, \citenamefont {Plastina},\ and\ \citenamefont
  {Paternostro}}]{lorenzo_geometrical_2013}%
  \BibitemOpen
  \bibfield  {author} {\bibinfo {author} {\bibfnamefont {S.}~\bibnamefont
  {Lorenzo}}, \bibinfo {author} {\bibfnamefont {F.}~\bibnamefont {Plastina}}, \
  and\ \bibinfo {author} {\bibfnamefont {M.}~\bibnamefont {Paternostro}},\
  }\href@noop {} {\bibfield  {journal} {\bibinfo  {journal} {Phys. Rev. A}\
  }\textbf {\bibinfo {volume} {88}},\ \bibinfo {pages} {020102} (\bibinfo
  {year} {2013})}\BibitemShut {NoStop}%
\bibitem [{\citenamefont {Yan}\ and\ \citenamefont
  {RuiXue}(2005)}]{yan_xu_annphyschem05}%
  \BibitemOpen
  \bibfield  {author} {\bibinfo {author} {\bibfnamefont {Y.}~\bibnamefont
  {Yan}}\ and\ \bibinfo {author} {\bibfnamefont {X.}~\bibnamefont {RuiXue}},\
  }\href@noop {} {\bibfield  {journal} {\bibinfo  {journal} {Annu. Rev. Phys.
  Chem.}\ }\textbf {\bibinfo {volume} {56}},\ \bibinfo {pages} {187} (\bibinfo
  {year} {2005})}\BibitemShut {NoStop}%
\bibitem [{\citenamefont {Hernandez-Gomez}\ \emph {et~al.}(2018)\citenamefont
  {Hernandez-Gomez}, \citenamefont {Poggiali}, \citenamefont {Cappellaro},\
  and\ \citenamefont {Fabbri}}]{capellaro_nvcenter_noise_2018}%
  \BibitemOpen
  \bibfield  {author} {\bibinfo {author} {\bibfnamefont {S.}~\bibnamefont
  {Hernandez-Gomez}}, \bibinfo {author} {\bibfnamefont {F.}~\bibnamefont
  {Poggiali}}, \bibinfo {author} {\bibfnamefont {P.}~\bibnamefont
  {Cappellaro}}, \ and\ \bibinfo {author} {\bibfnamefont {N.}~\bibnamefont
  {Fabbri}},\ }\href@noop {} {\  (\bibinfo {year} {2018})},\ \Eprint
  {http://arxiv.org/abs/1808.08222} {arXiv:1808.08222 [quant-ph]} \BibitemShut
  {NoStop}%
\bibitem [{\citenamefont {Pokharel}\ \emph {et~al.}(2018)\citenamefont
  {Pokharel}, \citenamefont {Anand}, \citenamefont {Fortman},\ and\
  \citenamefont {Lidar}}]{pokharel_demonstration_2018}%
  \BibitemOpen
  \bibfield  {author} {\bibinfo {author} {\bibfnamefont {B.}~\bibnamefont
  {Pokharel}}, \bibinfo {author} {\bibfnamefont {N.}~\bibnamefont {Anand}},
  \bibinfo {author} {\bibfnamefont {B.}~\bibnamefont {Fortman}}, \ and\
  \bibinfo {author} {\bibfnamefont {D.}~\bibnamefont {Lidar}},\ }\href@noop {}
  {\bibfield  {journal} {\bibinfo  {journal} {Phys. Rev. Lett.}\ }\textbf
  {\bibinfo {volume} {121}},\ \bibinfo {pages} {220502} (\bibinfo {year}
  {2018})}\BibitemShut {NoStop}%
\bibitem [{\citenamefont {Clerk}\ \emph {et~al.}(2010)\citenamefont {Clerk},
  \citenamefont {Devoret}, \citenamefont {Girvin}, \citenamefont {Marquardt},\
  and\ \citenamefont {Schoelkopf}}]{clerk_devoret_rmp10}%
  \BibitemOpen
  \bibfield  {author} {\bibinfo {author} {\bibfnamefont {A.~A.}\ \bibnamefont
  {Clerk}}, \bibinfo {author} {\bibfnamefont {M.~H.}\ \bibnamefont {Devoret}},
  \bibinfo {author} {\bibfnamefont {S.~M.}\ \bibnamefont {Girvin}}, \bibinfo
  {author} {\bibfnamefont {F.}~\bibnamefont {Marquardt}}, \ and\ \bibinfo
  {author} {\bibfnamefont {R.~J.}\ \bibnamefont {Schoelkopf}},\ }\href@noop {}
  {\bibfield  {journal} {\bibinfo  {journal} {Rev. Mod. Phys.}\ }\textbf
  {\bibinfo {volume} {82}},\ \bibinfo {pages} {1155} (\bibinfo {year}
  {2010})}\BibitemShut {NoStop}%
\bibitem [{\citenamefont {Shen}\ \emph {et~al.}(2017)\citenamefont {Shen},
  \citenamefont {Noh}, \citenamefont {Albert}, \citenamefont {Krastanov},
  \citenamefont {Devoret}, \citenamefont {Schoelkopf}, \citenamefont {Girvin},\
  and\ \citenamefont {Jiang}}]{shen_noh_prb17}%
  \BibitemOpen
  \bibfield  {author} {\bibinfo {author} {\bibfnamefont {C.}~\bibnamefont
  {Shen}}, \bibinfo {author} {\bibfnamefont {K.}~\bibnamefont {Noh}}, \bibinfo
  {author} {\bibfnamefont {V.~V.}\ \bibnamefont {Albert}}, \bibinfo {author}
  {\bibfnamefont {S.}~\bibnamefont {Krastanov}}, \bibinfo {author}
  {\bibfnamefont {M.~H.}\ \bibnamefont {Devoret}}, \bibinfo {author}
  {\bibfnamefont {R.~J.}\ \bibnamefont {Schoelkopf}}, \bibinfo {author}
  {\bibfnamefont {S.}~\bibnamefont {Girvin}}, \ and\ \bibinfo {author}
  {\bibfnamefont {L.}~\bibnamefont {Jiang}},\ }\href@noop {} {\bibfield
  {journal} {\bibinfo  {journal} {Phys. Rev. B}\ }\textbf {\bibinfo {volume}
  {95}},\ \bibinfo {pages} {134501} (\bibinfo {year} {2017})}\BibitemShut
  {NoStop}%
\bibitem [{\citenamefont {Caruso}\ \emph {et~al.}(2014)\citenamefont {Caruso},
  \citenamefont {Giovannetti}, \citenamefont {Lupo},\ and\ \citenamefont
  {Mancini}}]{caruso_rmp14}%
  \BibitemOpen
  \bibfield  {author} {\bibinfo {author} {\bibfnamefont {F.}~\bibnamefont
  {Caruso}}, \bibinfo {author} {\bibfnamefont {V.}~\bibnamefont {Giovannetti}},
  \bibinfo {author} {\bibfnamefont {C.}~\bibnamefont {Lupo}}, \ and\ \bibinfo
  {author} {\bibfnamefont {S.}~\bibnamefont {Mancini}},\ }\href@noop {}
  {\bibfield  {journal} {\bibinfo  {journal} {Rev. Mod. Phys.}\ }\textbf
  {\bibinfo {volume} {86}},\ \bibinfo {pages} {1203} (\bibinfo {year}
  {2014})}\BibitemShut {NoStop}%
\bibitem [{\citenamefont {Filippov}\ \emph {et~al.}(2017)\citenamefont
  {Filippov}, \citenamefont {Magadov},\ and\ \citenamefont
  {Jivulescu}}]{filippov_sergey_njp17}%
  \BibitemOpen
  \bibfield  {author} {\bibinfo {author} {\bibfnamefont {S.~N.}\ \bibnamefont
  {Filippov}}, \bibinfo {author} {\bibfnamefont {K.~Y.}\ \bibnamefont
  {Magadov}}, \ and\ \bibinfo {author} {\bibfnamefont {M.~A.}\ \bibnamefont
  {Jivulescu}},\ }\href@noop {} {\bibfield  {journal} {\bibinfo  {journal} {New
  J. Phys.}\ }\textbf {\bibinfo {volume} {19}},\ \bibinfo {pages} {083010}
  (\bibinfo {year} {2017})}\BibitemShut {NoStop}%
\end{thebibliography}


\end{document}